\documentclass[acmsmall]{acmart}

\usepackage{enumerate}
\usepackage{subcaption}
\usepackage{siunitx}
\usepackage{amsmath}
\usepackage{algorithm}
\usepackage{algorithmic}
\usepackage{stfloats}
\usepackage{caption}
\usepackage{multirow}
\usepackage{xcolor}
\usepackage{amsfonts}
\usepackage{soul,color}
\usepackage{graphicx}
\usepackage{lipsum}
\usepackage{xurl}
\usepackage{nccmath}
\usepackage{wrapfig}
\usepackage{booktabs} 
\usepackage{makecell} 
\usepackage{array}
\usepackage{float}
\usepackage{enumitem}
\usepackage{hyperref}

\newcolumntype{?}{!{\vrule width 1pt}}
\newcommand\rtt[1]{\rotatebox[origin=c]{90}{#1}}
\newcommand\tcb[1]{\textbf{\textcolor{red}{#1}}}
\makeatletter
\def\thickhline{%
  \noalign{\ifnum0=`}\fi\hrule \@height \thickarrayrulewidth \futurelet
   \reserved@a\@xthickhline}
\def\@xthickhline{\ifx\reserved@a\thickhline
               \vskip\doublerulesep
               \vskip-\thickarrayrulewidth
             \fi
      \ifnum0=`{\fi}}
\makeatother
\newlength{\thickarrayrulewidth}
\newlength{\Oldarrayrulewidth}
\newcommand{\Cline}[2]{%
  \noalign{\global\setlength{\Oldarrayrulewidth}{\arrayrulewidth}}%
  \noalign{\global\setlength{\arrayrulewidth}{#1}}\cline{#2}%
  \noalign{\global\setlength{\arrayrulewidth}{\Oldarrayrulewidth}}}
\newcolumntype{L}{>{\centering\arraybackslash}m{0.8cm}}
\newcolumntype{M}{>{\arraybackslash}m{7.5cm}}

\newcolumntype{C}{>{\centering\arraybackslash}m{1.9cm}}
\newcolumntype{D}{>{\arraybackslash}m{11cm}}

\AtBeginDocument{%
  \providecommand\BibTeX{{%
    \normalfont B\kern-0.5em{\scshape i\kern-0.25em b}\kern-0.8em\TeX}}}

\setcopyright{acmcopyright}
\copyrightyear{2018}
\acmYear{2018}
\acmDOI{10.1145/1122445.1122456}

\acmJournal{JETC}
\acmVolume{37}
\acmNumber{4}
\acmArticle{111}
\acmMonth{8}

\begin{document}

\title{A Non-invasive Technique to Detect Authentic/Counterfeit SRAM Chips}

\author{B. M. S. Bahar Talukder}
\email{bbaha007@fiu.edu}
\orcid{0000-0001-6388-0509}
\author{Farah Ferdaus}
\email{fferd006@fiu.edu}
\orcid{0000-0001-5510-6193}
\author{Md Tauhidur Rahman}
\email{mdtrahma@fiu.edu}
\orcid{0000-0002-0010-6388}
\affiliation{%
  \institution{Department of ECE, Florida International University}
  \streetaddress{10555 West Flagler Street}
  \city{Miami}
  \state{Florida}
  \postcode{33174}
  \country{USA}
}

\renewcommand{\shortauthors}{Bahar Talukder, et al.}

\begin{abstract}
Many commercially available memory chips are fabricated worldwide in untrusted facilities. Therefore, a counterfeit memory chip can easily enter into the supply chain in different formats. Deploying these counterfeit memory chips into an electronic system can severely affect security and reliability domains because of their sub-standard quality, poor performance, and shorter lifespan. Therefore, a proper solution is required to identify counterfeit memory chips before deploying them in mission-, safety-, and security-critical systems. However, a single solution to prevent counterfeiting is challenging due to the diversity of counterfeit types, sources, and refinement techniques. Besides, the chips can pass initial testing and still fail while being used in the system. Furthermore, existing solutions focus on detecting a single counterfeit type (e.g., detecting recycled memory chips). This work proposes a framework that detects major counterfeit static random-access memory (SRAM) types by attesting/identifying the origin of the manufacturer. The proposed technique generates a single signature for a manufacturer and does not require any exhaustive registration/authentication process. We validate our proposed technique using 345 SRAM chips produced by major manufacturers. The silicon results show that the test scores ($F_{1}$ score) of our proposed technique of identifying memory manufacturer and part-number are 93\% and 71\%, respectively.
\end{abstract}



\maketitle

\section{Introduction}\label{sec:intro}
\begin{figure}[ht!]
  \centering
  \captionsetup{justification=centering, margin= 0.0cm}
  \includegraphics[width=0.6\textwidth, trim=0 0cm 9.2cm 13cm, clip]{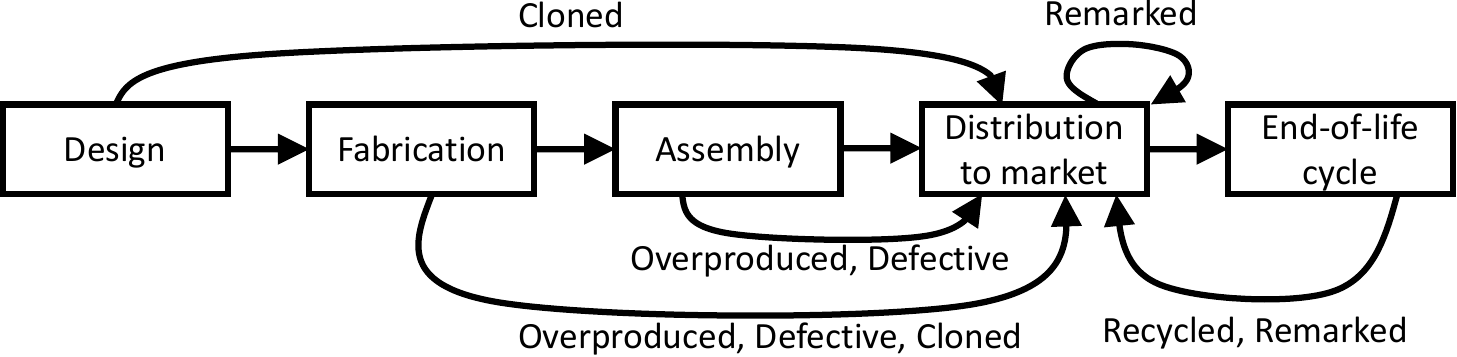} 
  \caption{Device counterfeiting at different stages of the horizontal semiconductor supply chain \cite{DRAM_Host:Talukder}.}
  \label{fig:hor_sup}
\end{figure}

With the globalization of the semiconductor supply chain and the growth of the semiconductor market value, counterfeit ICs have become an established threat to the semiconductor community. In the modern horizontal semiconductor supply chain, multiple facility centers are involved in the manufacturing process and facilitate different stages of chip production. In this supply chain model (Fig. \ref{fig:hor_sup}), a chip may be designed in one place and fabricated in different places. Because of these traveling IPs (intellectual properties) in different formats, the device can be easily counterfeited in many different ways and may easily get introduced to the consumer market. Recent studies show that the global market share of counterfeit integrated circuits (ICs) worth \$169 billion \cite{CounterfeitIC:UGuin, SCARE_GUO}, and $\sim$17\% of those counterfeited ICs are memory chips \cite{CounterfeitIC:UGuin, SCARE_GUO, Forte:CHES}. 
Moreover, another $\sim$28\% of the counterfeit chips are contributed by programmable logic (CPLDs, FPGAs), microcontrollers, and microprocessors \cite{Forte:CHES}. However, most of the modern programmable logic and microcontrollers/microprocessors are integrated with memory (e.g., BRAM in FPGA, cache in microcontrollers/microprocessors). Therefore, identifying counterfeit memory should be able to capture the majority of these counterfeit programmable logic and microcontrollers/microprocessors. 
Counterfeit chips are classified into the following major categories \cite{CounterfeitIC:UGuin}: (i) recycled, (ii) remarked or forged documentation, (iii) tampered (iv) cloned, (v) reverse-engineered, (vi) out-of-spec/defective, and (vii) overproduced. Table \ref{tab:countefeitDef} shows various types of counterfeitings along with examples \cite{Forte:CHES, CounterfeitIC:UGuin}. A counterfeit chip suffers inferior quality and, therefore, can impact the safety, security, and reliability of a system \cite{ARMEDSERVICES:Counterfeit}. For example, Russia's recent Fobos-Grunt mission to Mars was canceled due to a counterfeit SRAM memory chip \cite{fobos_Grunt}. Unfortunately, identifying counterfeit chips is tricky as they can pass the initial functional test but may fail prematurely due to their lower life expectancy than authentic chips.

\begin{table}[ht!]
\caption{Different types of counterfeiting \cite{Forte:CHES, CounterfeitIC:UGuin}.}
\begin{tabular}{|C|D|}
\hline
Counterfeit Types &
  Definition\\ \hline
Recycled &
  Recycling chips from old PCB and selling them as new. In a more sophisticated recycling process, the plastic/ceramic encapsulation of the chip die is removed and repackaged to make its appearance new. Recycled chips shares $>$80\% of the counterfeit market \cite{Forte:CHES}.\\ \hline
Remarked &
  Inferior quality chips are marked as the superior one. \\ \hline
Forged documented &
  Faking the chip documentation (e.g., faking safety and security certification). \\ \hline
Reverse   Engineered &
  Recover the functional netlist from the chip by an electro-chemical process. Counterfeiter may use this netlist to avoid the R\&D cost. \\ \hline
Cloned &
  The untrusted fabrication facility can copy the chip design (netlist, GDSII); later, they can produce unauthorized chips. \\ \hline
Overproduced &
  Untrusted fabrication facility can produce and market chips outside of the contract, i.e., without authorization of the IP (intellectual  properties) owner. \\ \hline
Tampered &
  Tampering the original chip design. For example, untrusted physical design house can insert hardware trojan in the netlist and create a security backdoor. \\ \hline
Out-of-spec/defective &
  Selling chips that are failed in the   functionality test (manufacturer name and part-number are removed and replaced with a superior one).\\ \hline
\end{tabular}
\label{tab:countefeitDef}
\end{table}

To combat the recent trend of increasing fake parts, the U.S. Government passed the National Defense Authorization Act (NDAA) in August 2018 \cite{NDAA}. Section 818 of this Act requires defense contractors to tighten supply chain traceability and parts procurement to minimize counterfeit risk \cite{Counter:Damage}. Researchers and industries have developed several techniques to detect and avoid counterfeit electronic components, such as physical inspections, imaging techniques, electrical testings, etc. \cite{CounterfeitIC:UGuin, SCARE_GUO, recycledSRAm:Guo, SRAMPUF:Xiao, SRAMPUF:Xu, Counterfeit:UGuin,Counterfeit:Goetz}. Unfortunately, most solutions focus on identifying a single type of counterfeit chips, e.g., detecting recycled memory chips \cite{SCARE_GUO, recycledSRAm:Guo,Recycled:Guin}. Furthermore, many of those techniques require either hardware modification, complex supply chain management, complex authentication schemes, or unique laboratory facilities \cite{Counterfeit:UGuin, CSST_Rahman, splitManufacturing, HWmetering,maisensor,guinsensor}. Hence, those are not suitable for low-cost memory chips.Deploying counterfeit chips into an electronic system can severely compromise system security and reliability because of their sub-standard quality, poor performance, and shorter life span. Unfortunately, identifying counterfeit chips is tricky as they can pass the initial functional test but may fail prematurely due to their lower life expectancy than authentic chips. 

Our recent studies \cite{DRAM_Host:Talukder} show that analyzing latency-based error patterns can capture manufacturers' information and DRAM module specifications. In this paper, we present a more generalized technique to detect and avoid major counterfeit SRAM types. In our proposed technique, we attest and identify the origin of SRAM chips (i.e., manufacturer and specification) by characterizing the start-up behavior of SRAM chips. Attesting and identifying memory manufacturers and specifications might be a powerful tool in avoiding the remarked, defective, tampered, and cloned memory chips. We find that the start-up behavior of SRAM chips varies from one manufacturer to another manufacturer and from one set of specifications to another set specifications because of intentional architectural/layout differences and the manufacturing process variations. Furthermore, we show that a similar analysis of SRAM start-up data can be used to identify recycled SRAM chips as the SRAM start-up behavior is directly correlated with its usage time. We also explore the robustness of our proposed technique and provide a guideline for practical implementation. The major contributions of this paper include:
\begin{itemize}[leftmargin=*, topsep=0pt,itemsep=-1ex,partopsep=1ex,parsep=1ex]
\item We have extracted a set of features from the start-up state of SRAM chips to capture the architectural, layout, and process variations. We found that our proposed set of features can be used to identify the memory manufacturer and part-number\footnote{A unique part-number is usually assigned to a group of electronic components that possess a similar set of specifications.}.
\item We have tested the robustness of our proposed method by varying operating temperature and testing platforms.
\item We have also compared the extracted features between the fresh and aged (recycled) chips. The practical aging state of SRAM memory has been emulated by stressing the memory chip under high-temperature and supply-voltage conditions. 
\item We have validated our proposed technique with the data collected from 345 commodity SRAM chips (produced by five major manufacturers).
\item We have provided a practical guideline to improve the accuracy of our proposed method with a realistic demonstration.
\end{itemize}

The rest of the paper is organized as follows- in Sec. \ref{sec:background-motivation}, we have briefly discussed SRAM structure, the aging effect on SRAM chips, and existing anti-counterfeit techniques. In Sec. \ref{sec:method}, we have proposed our method of extracting an appropriate set of features from SRAM start-up data. In Sec. \ref{sec:result}, we have presented our experimental results and analyzed them. We have highlighted the scope and limitations of this work along with the future work in Sec. \ref{sec:discussion}. We have concluded our work in Sec. \ref{sec:conclusion}.

\section{Background and Motivations}\label{sec:background-motivation}
This section briefly describes SRAM architecture, the aging effect on SRAM cells, and the existing approach to detect counterfeit memory chips. 

\subsection{SRAM, Process Variations, and Aging}\label{subsec:cellArch}
\begin{figure}[ht!]
  \centering
  \captionsetup{justification=centering, margin= 0.0cm}
  \includegraphics[width=0.35\textwidth]{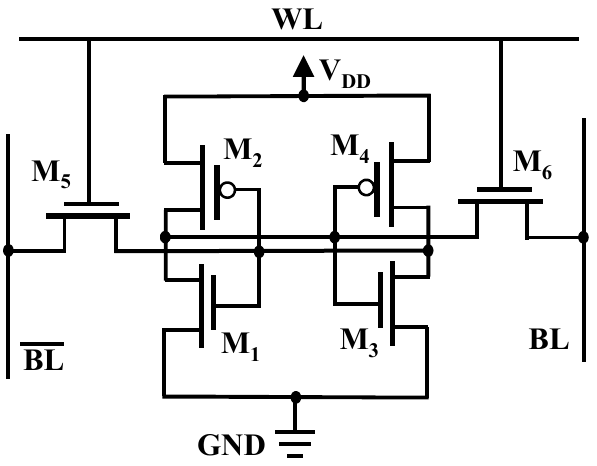} 
  \caption{SRAM cell structure.}
  \label{fig:SRAMcell}
\end{figure}

SRAM cell, volatile memory that stores one-bit data, consists of two cross-coupled inverters and two access transistors (see Fig. \ref{fig:SRAMcell}) \cite{SCARE_GUO, starupModel:Cortez}. The cross-coupled inverters are symmetrically laid out to maximize the static noise margin ($SNM$) \cite{SCARE_GUO, starupModel:Cortez}. $SNM$ is defined as the maximum allowable noise that can tolerate an SRAM cell without flipping its value \cite{SNM:Mukherjee}. However, the inevitable random dopant fluctuation (RDF) effect leads to threshold voltage variation and introduces asymmetricity between SRAM inverters \cite{Heterogeneous:Kwon}. Therefore, during power-up, these two inverters race each other and settle to ``1" or ``0" \cite{SCARE_GUO, starupModel:Cortez}. A significant difference between inverters' strength generates a strong ``0" or a strong ``1". On the other hand, a smaller difference between the two inverters generates weak ``0" or weak ``1". Furthermore, the smallest difference between the two inverters creates a noisy start-up value.

Moreover, two well-known phenomena, negative and positive bias temperature instability (NBTI, PBTI), can also cause transistor threshold voltage to shift \cite{Recycled:Guin}. NBTI and PBTI are the direct consequence of transistor aging \cite{Recycled:Guin}. Previous research suggests that the SNM of the SRAM decreases by $>$9\% within three years of usages \cite{NBTI:Park}.
 
\subsection{Memory Supply Chain Vulnerabilities} \label{sec:supplyChain}
Globalization of the semiconductor supply chain has allowed worldwide fabrication of authentic and counterfeit chips \cite{CounterfeitIC:UGuin, PVal:Basak}. In an established global semiconductor supply chain, several untrusted parties (foundry, assembly, third-party IPs, etc.) are involved, any of whom can pirate IP (intellectual property), insert hardware trojan, and/or include recycled, re-marked, overproduced, out-of-spec/defective, cloned, and forge-documented chips \cite{PVal:Basak, DRAM_Host:Talukder}. In this global supply chain, the IC or memory manufacturer can (\textbf{i}) fabricate all memory chips or ICs in a single manufacturer-owned foundry or (\textbf{ii}) can send the Graphic Design System (GDSII) file (a file format that contains the final-layout information) to several foundries of their own (but in different geolocation) or third-party foundries to save on the cost per unit or to meet the target timeline \cite{DRAM_Host:Talukder}. A counterfeiter can sell fake memory chips as authentic ones, recycled or used memory chips as new ones through repackaging, low-quality chips as high-grade ones by mislabeling, and defective or out-of-spec chips without the manufacturer's consent. An adversary in an untrusted foundry can insert a hardware trojan in the form of addition, deletion, or modification of memory cell, memory array, or peripheral logic \cite{memTrojan}, which changes the memory layout/architecture \cite{memTrojan}.

\subsection{Existing Countermeasures and Limitations} \label{sec:limitaion}
There have been several techniques to detect counterfeit chips. Some existing approaches rely on generating signatures from individual chips \cite{SCARE_GUO, starupModel:Cortez, Powerup:Holcomb}. One chip can not be cloned to another chip because of the signatures' uniqueness due to the process variation. These memory signatures vary from chip to chip, even if they are fabricated in the same silicon wafer. Such signatures are well-known as physical unclonable functions (PUF). The signature from the individual chip is collected and stored in the database during the registration process. During authentication, signatures are collected from the memory under test (MUT) and compared with the database. A device is considered authentic if its signature matches with the expected stored signature. The database can store memory fingerprints from a single measurement. However, the memory signatures are noisy and can be affected by operating conditions. Recently, Guo \textit{et al.} propose measuring memory signature at both room temperature and high temperature to compute a more robust signature \cite{SCARE_GUO}. However, the PUF-based method suffers from several limitations:
\begin{itemize}[leftmargin=*, topsep=0pt,itemsep=-1ex,partopsep=1ex,parsep=1ex]
\item \textbf{Bit-aliasing:} Bit-aliasing measures the uniqueness and correlation among signatures (PUFs) \cite{SRAMPUF:tauhid}. It quantifies the distribution of ``0" or ``1" on a specific memory cell. The bit aliasing can be quantified with Eq. \ref{eqn:bit-alias}.
\begin{equation} \label{eqn:bit-alias}
B_{i} = \frac{1}{N}\sum_{p=1}^{N} {R_{p}^{i,l}}
\end{equation}
Here, \textit{N} is the total number of devices needed to be identified uniquely, where each device is equipped with an \textit{l}-bit PUF. ${R_{p}^{i,l}}$ is the PUF response recorded from the \textit{i}th bit of the \textit{p}th PUF. In an ideal case, the mean occurrence of logic ``0" or ``1" from a specific bit location should be 50\% (i.e., bit-aliasing should be 50\%). The ideal bit-aliasing of 50\% minimizes the number of bits required to identify all devices uniquely. For example, in an ideal case, to identify 4 SRAM chips uniquely, we need only a 2-bit PUF response (i.e., ``00", ``01", ``10", and ``11"). In such a case, the occurrence of ``1" or ``0" on the first-bit position or the second-bit position is 50\%. However, if the average occurrence deviates from 50\%, we might need more than 2-bits to authenticate those four SRAM chips. In practice, the bit-aliasing always deviates from 50\% and requires more bits than it needed theoretically.

\item \textbf{Exhaustive registration process:} The signature-based chip authentication requires registering each memory chip before distributing them in the market. This extra step of registration increases both cost and lead time to market.

\item  \textbf{Robustness:} Device signature also may vary depending on the operating condition. A slight variation on temperature or operating voltage might alter the device characteristics and flip some bits on the device signature. Although different Error Correcting Codes (ECC) \cite{AROPUF:tauhid, PUF:Bhargava} are proposed as a solution; however, the ECC overhead increases quadratically with the number of errors \cite{AROPUF:tauhid}.
\end{itemize}

Other countermeasures such as SST, hardware metering, blockchain-based traceability, split manufacturing, IC camouflaging, Electronic Chip ID (ECID), On-chip sensor, DNA marking, etc., might be used to prevent counterfeiters \cite{CounterfeitIC:UGuin, Forte:CHES, CSST_Rahman, splitManufacturing, HWmetering, maisensor, guinsensor, onChipSensor:Zhang, onChipSensor:He, Huang:CounterfeitIC, hwmeter:Alkabani, CDIR:Zhang, DNA:Hayward, RFID, Kundu, Rajendran:camouflage}; however, these techniques suffer from different drawbacks. For example, SST and hardware metering techniques provide control over post-fabrication, but it requires a change in traditional fabrication flow. Furthermore, this technique requires exhaustive communication between the foundry and the manufacturer. On the other hand, ECID tags each chip with a unique ID by adding a one-time programmable (OTP) memory. Nevertheless, this method is not suitable for all kinds of chips. For an SRAM chip, the overhead of adding an extra memory component will be very high. With an on-chip sensor, each chip is equipped with an additional hardware component, which modifies its properties due to aging. These properties can be used to detect recycled chips. However, on-chip sensor-based countermeasures need additional hardware overhead and are not feasible for inexpensive systems. In DNA marking, each memory component is marked with a unique DNA sequence. DNA marking suffers from impracticality as it requires a complex authentication scheme. Other techniques, such as blockchain-based traceability, split manufacturing, IC camouflaging, etc., require modified fabrication flow or design techniques that are not suitable for low-cost memory chips.

Physical inspection-based schemes \cite{CounterfeitIC:UGuin, foundaryIdentification:Wendt, ML:Ahmadi, forensic:Helinski}, such as X-Ray imaging and scanning electron, can detect counterfeit/recycled chips.  However, these techniques require expensive equipment and not viable for inexpensive chips. Moreover, Expensive equipment and complex authentication schemes are also not suitable for general users who want to verify their purchased products' authenticity.

This paper proposed a technique to detect counterfeit SRAM chips that do not suffer from the above limitations.

\section{Proposed Method}\label{sec:method}
By analyzing the internal signatures of the SRAM memory chips, our proposed technique will identify major types of counterfeit chips by- (\textbf{i}) attesting  the origin of the memory chip manufacturer and the specification (i.e., the part-number) of each memory chip and (\textbf{ii}) detecting recycled memory chips. This section describes sources of distinguishable factors, unique features that isolate one part-number with another or identify the same part-number, and our proposed framework.
\subsection{Sources of Distinguishable Factors} \label{sec:distFactor}
Our proposed technique relies on the fact that SRAM chips of different specifications differ with architectural, layout, and process parameters, which leads to unique GDSII. All these factors can be used to generate a unique signature from each group of SRAMs.
\begin{itemize}[leftmargin=*, topsep=0pt,itemsep=-1ex,partopsep=1ex,parsep=1ex] 
    \item \textbf{Architectural variations:} Manufactures may optimize the SRAM structure in different ways to support the requirement \cite{SRAM:Rollini, SRAM:Asthana, SRAM:Chuang, SRAM:Chnag}. Among different structures, the symmetric 6-Transistor (6T) SRAM structure is the most common one (Fig. \ref{fig:SRAMcell}) for on-chip SRAM array (e.g., processors cache). 4T SRAM cells are also common for off-chip SRAM memory. However, 4T SRAM chips can not be implemented on-chip as they need different technology and complex process. Theoretically, the symmetric structure of SRAM cells should produce a uniform distribution of logic ``0" and logic ``1". On the other hand, to suppress the noise (e.g., read disturbance, half-select disturbance, etc.), other SRAM architecture such as 5T, asymmetric 6T, 7T, 8T, 9T, 10T structure is also available \cite{SRAM:Rollini, SRAM:Asthana, SRAM:Chuang, SRAM:Chnag}. However, due to these configurations' asymmetric structure, each SRAM cell on the memory array may be biased to a specific logic at start-up. Furthermore, to reduce the bitline noise, the bitlines are often twisted in different configurations \cite{SRAM:Weste}. The difference in bitline configuration also may affect the start-up logic locality.
    \item \textbf{Layout Variations:} The layout variation in SRAM cell structure may also cause a variation in start-up characteristics. For example, Apostolidis \textit{et al.} \cite{SRAM:Apostolidis} reported six different layout designs for symmetric 6T SRAM structure, and each of them has different pros and cons. For example, they have different power utilization, delay, and noise characteristics. In addition to this, some implementing and resource constrain may introduce some asymmetric nature in memory cells, leading to slight bias to a specific logic at device start-up. For example, using multiple metal layers may introduce unmatched wiring between the inverter pair. Moreover, the difference in CAD tools' configurations  may also introduce variations in memory layout.
    \item \textbf{Process variations:} The intrinsic process variation can be either random or systematic \cite{processVar:Cao, processVar:kuhn}. The random process variation can be considered the noise and can be varied among the chips fabricated in a single wafer. However, the systematic process variation can be introduced by the quality of the fabrication plant (foundry), microarchitectural locality, and pattern. For the symmetric layout design of the symmetric 6T SRAM cell, the layout of one inverter is the mirror to the other one. However, the fabrication plant may have different set of rules for mirrored patterns \cite{VLSI-SoC}. Hence, a mirrored layout may be reffed as a different pattern when fabricated. Hence, even with the perfectly symmetric layout design, the two coupled inverters may have slightly different characteristics after fabrication. Additionally, a recent study shows that the founder-dependent channel length and threshold voltage variations impact the IC delay characteristics \cite{foundaryIdentification:Wendt}.
    \item \textbf{IC packaging:} Chip die is encapsulated inside a protected ``package" to prevent corrosion and physical damage. The difference in IC packaging may also alter some device characteristics. Usually, manufacturers introduce different kinds of die packaging and wire bonding to trade-off among cost, noise immunity, and supporting different operating conditions \cite{package:Khazaka}. The impact of die packaging is minimal compared to other factors, as the IC packaging only impacts external influences (such as noise induced by external temperature). However, the die packaging may influence some device characteristics and, therefore, may impact some of our selected features (see Sec. \ref{sec:FeatSelect}). For example, the chips with gold/copper wire bonding should be more robust against external temperature variation than those with aluminum wire bonding. Therefore, the noise magnitude in SRAM start-up data should be smaller with gold/copper wire bonding.
    \item \textbf{Aging:} Usually, the SRAM signature (PUF) can be characterized by $PSNM_{noise}$ (PUF $SNM$ noise) \cite{starupModel:Cortez, Masoumian:PSNM}. The $PSNM_{noise}$ measures how easily an SRAM cell can be initiated to logic ``0" or ``1". A larger value of $PSNM_{noise}$ ensures more robust SRAM signatures. However, the $PSNM_{noise}$ heavily depends on SRAM transistors' threshold voltage \cite{starupModel:Cortez}. Hence, the SRAM $PSNM_{noise}$ can be changed over its usages (see Sec. \ref{subsec:cellArch}) due to the change in its transistors' threshold voltage.
    
    Depending on SRAM usage data pattern, the change in $PSNM_{noise}$ can affect the SRAM start-up signature: (\textbf{i}) a noisy signature bit might get biased to ``0" or ``1", (\textbf{ii}) a weak ``0" or ``1" might become strong ``0" or ``1", (\textbf{iii}) a stable signature bit can be flipped (stable ``0" to  stable ``1" or stable ``1" to stable ``0"), and (\textbf{iv}) a stable signature bit can become a noisy one. Hence, the change in $PSNM_{noise}$ will affect the overall distribution of logic ``0" and ``1" on SRAM signature. The first three factors will increase the total number of stable signature bits; whereas, the fourth factor will produce more noisy signature bits. However, the cumulative impact of the first three factors dominates the fourth factor. Hence, the total number of noisy signature bits will reduce with device usage (which does not indicate the PUF will be more robust with aging \cite{SRAMPUF:tauhid}). Minimizing the mismatch between two inverters can strengthen the impact of the fourth factor, which is difficult to achieve. The equalization of transistors' threshold voltage requires a calculative usage data pattern during the entire chip lifetime \cite{NBTI:Park}.
    
    In an ideal case, the percentage of 0’s or 1’s should be identical in a new symmetric SRAM chip. One of the recent methods suggests that the skewed distribution of 0’s and 1’s at power-up state can be used to detect recycled SRAM memory \cite{Recycled:Guin}. With a typical usage pattern, an SRAM cell experiences more logic ``0" bits than the logic ``1" bits \cite{Wei:NibRemap}. Such usages pattern creates more stress on ``M4" pMOS (Fig. \ref{fig:SRAMcell}). Hence, over time, the threshold voltage difference of ``M4" and ``M2" PMOS increases due to the NBTI effect and causes the SRAM cell to be biased with ``1" at power-up state. Note that this method of detecting recycle memory is a special case of our proposed technique.
\end{itemize}

\subsection{Assumptions} \label{sec:Assumption}
Our proposed technique extracts a set of features from memory signatures and uses them to train a statistical model and identify manufacturer/part-number. Although our method uses a simple authentication protocol, we make the following set of assumptions which are practical for most usage scenarios.
\begin{itemize}[leftmargin=*, topsep=0pt,itemsep=-1ex,partopsep=1ex,parsep=1ex]
    \item \textbf{Defining features:} Manufacturers/trusted third-parties are responsible for defining a set of features that defines their product best. Prior knowledge of memory architecture might enable them to define a better set of features.
    \item \textbf{Feature extraction:} The feature extraction process should be independent and straightforward enough to be extracted on the user’s system; hence, it relaxes the requirement of any special tool or environment requirement. We also assume that the user does not have any knowledge of memory architecture; only general information available from manufacturers should enable a user to extract the features.
    \item \textbf{Memory Class:} Two memories are from separate classes if they have a different manufacturer and/or a different set of specifications (i.e., speed, size, temperature range, power rating, data-width, die package, die generation, etc.). A change in specification and/or manufacturer lead to different GDSII and/or packaging; hence, it will impact start-up data, as discussed in Sec. \ref{sec:method}. Although a manufacturer may send the same GDSII to multiple fabrication facilities, we assume that fabrication plants with the same GDSII maintain the same design rule to keep uniformity. We also assume that a manufacturer may produce memories with a different specification but with the same set of fabrication plants or design memories with a slight change in specifications (for example, only change in the die package). In such a case, these memories may have two sets of features with subtle variation, which leads to a complex classification problem to identify the memory correctly.
    \item \textbf{Classification:} Classifying memory (authentic vs. counterfeit) can be done in either manufacturer end or consumer end, depending on the application. For example, if the manufacturer is reluctant to release the statistical model publicly, it might ask for the features from memory under test (MUT) to verify the authenticity. On the other hand, to reduce the communication overhead and complexly, the manufacturer may release the statistical model publicly, and the MUT can be verified on the user's system. 
\end{itemize}

\subsection{Feature Selection}\label{sec:FeatSelect}
The accuracy and efficiency of any machine learning algorithm heavily rely on the features that are used for the algorithm. Hence, in this step, we proposed a set of SRAM start-up-based features that can effectively capture the architectural, layout, and process variations. A good feature should obtain (\textbf{i}) the similarities of chips with the same specification and (\textbf{ii}) the discrepancy between chips manufactured with different specifications. 

In our proposed method, we collected $n$ sets of start-up data ($\{D_{1}, D_{2}, \reflectbox{$\cdots$}, D_{n}\}$) from each SRAM chip. We constructed a unified data, $D$, based on majority voting\footnote{In the majority voting technique, each signature bit is sampled multiple times, and the value of that signature bit is assigned as the majority of the samples \cite{MajVoting}.} cast by $\{D_{1}, D_{2}, \reflectbox{$\cdots$}, D_{n}\}$. SRAM memory cells are generally arranged in a 2-D array of size $r\times c$ ($r=  number\ of\ rows$ and $c= number\ of\ columns$). If each word of a SRAM chip consists of $w_{l}$ bit data, then, for simplicity, we can assume that there is a total of $w_l$ 2-D array of single bit contributing 1-bit data to each data word. So, the data $D$ should be 3-D data of size $r\times c \times w_{l}$. However, to reduce the complexity, we rearrange the whole data in a 2-D array of size $n_{w}\times w_{l}$ ($= dim(D)$), where $n_{w} = r\times c$ is the number of words in the memory. Now we extract the following seven features from the start-up data $D$ \cite{DRAM_Host:Talukder}:

\begin{itemize}[leftmargin=*, topsep=0pt,itemsep=-1ex,partopsep=1ex,parsep=1ex]
\item \textbf{\textit{Feature 1 ($\Phi_1$)}:} This feature quantifies the ``cell biasness" by counting the number of logic ``1" bits in the start-up data. 
The evaluation of $\Phi_1$ is illustrated in Fig. \ref{fig:toyExample}. In this example, we presented start-up data from an 4$\times$8 ($n_{w}\times w_{l}$) SRAM chip containing four 8-bit words. In this figure, 16 bits contain logic ``1" out of 32 bits. Hence, according to our definition, $\Phi_1 = 16/32 = 0.5$.
Cell bias qualitatively measures the asymmetricity of the cross-coupled inverters (see Sec. \ref{subsec:cellArch}). For an ideally symmetric SRAM cell structure, this value should be 0.5 (i.e., no ``cell bais"). However, in practice, this value is usually deviated from 0.5 because of the different variations discussed previously (see Sec. \ref{sec:distFactor}).
\begin{figure}[ht!]
  \centering
  \includegraphics[width=0.5\textwidth, trim=0 11.1cm 17.4cm 0, clip]{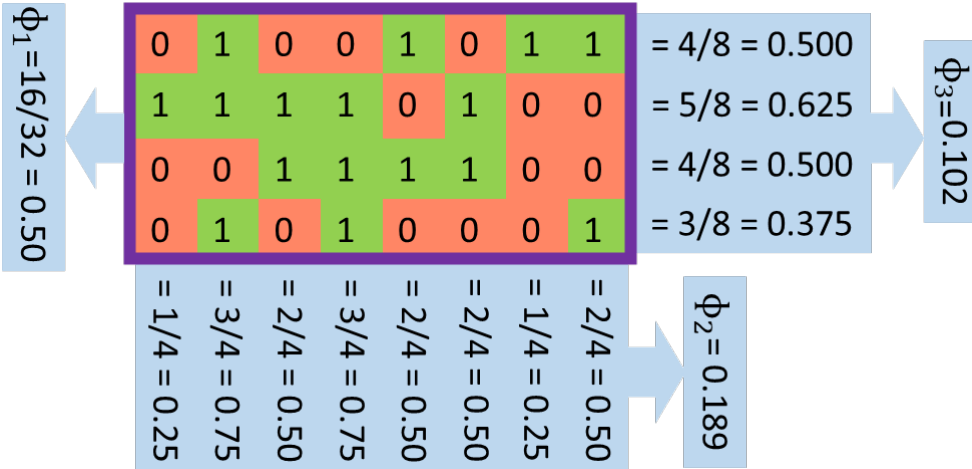} 
  \caption{Illustration of $\Phi_1$, $\Phi_2$, and $\Phi_3$ (4$\times$8 SRAM).}
  \label{fig:toyExample}
\end{figure}
\item \textbf{\textit{Feature 2 ($\Phi_2$)}:} An SRAM chip of word size $w_{l}$ can be assumed as a series of $w_{l}$ 2-D SRAM arrays. We counted a fraction of ``1" from each 2-D array for this feature and took the standard deviation as the feature $\Phi_2$. If each of the 2-D arrays follows similar data distribution, and the $\Phi_2$ should be close to 0. In Fig. \ref{fig:toyExample}, each 2-D array is rearranged in a single-dimensional vector for visualization purposes and presented along each column. Now, to evaluate $\Phi_2$, we computed the fraction of logic ``1" along each column, and then standard deviation is calculated using those values. $\Phi_2$ can capture different physical properties of the SRAM chips. For example, if the area constraint is too tight, all 2-D memory arrays can be located in close proximity or may be fused together. In that case, they may have a smaller difference in logic distribution due to smaller process variations.
\item \textbf{\textit{Feature 3 ($\Phi_3$)}:} The fraction of logic bit ``1" is counted in each word of data $D$; then, the standard deviation of those values was taken as the feature $\Phi_3$. $\Phi_3$ is also illustrated using Fig. \ref{fig:toyExample}. In this figure, we first calculated the fraction of logic ``1" from each word (along the row), and then $\Phi_3$ is estimated by computing the standard deviation of those values.
In an ideal case, the distribution of logic bit ``1" from each data word should be normally distributed with a mean of 50\%. Our experimental results demonstrate that the mean is close to $\Phi_1$. However, the standard deviation of distribution may vary from chip to chip depending on  memory specification (i.e., for some memory chips, the distribution can be flatter than other chips of different specifications). $\Phi_3$ quantifies the symmetricity of the SRAM cell array. For example, each SRAM cell might experience different systematic process variations due to the local layout patterns\footnote{Local layout patterns might be different from one cell to another, e.g., memory cells near the sense amplifier vs. memory cells at the middle of the SRAM array.}; hence, data words from different address locations might experience different logic distribution at start-up. A larger variation on local logic distribution will result in a larger value of $\Phi_3$.
\item \textbf{\textit{Feature 4 ($\Phi_4$)}:} The compression ratio ($r$, where, $r\geq1$) of the start-up data is selected as one of the features. A start-up data with regular patterns have larger data redundancy and can be significantly compressed without any information loss. However, start-up data with randomly distributed zeros and ones can be squeezed very little and causes a smaller value of compression ratio (closer to 1). $\Phi_4$ can capture the impact of the random process variation on SRAM chips. The compression ratio is defined as Eq. \ref{eqn:compRatio}.

\begin{equation} \label{eqn:compRatio}
r = \frac{\mathcal{S}_{u}}{\mathcal{S}_{c}}
\end{equation}
Where,
\begin{align*} 
& \mathcal{S}_{u} = size\ of\ the\ uncompressed\ data\\
& \mathcal{S}_{c} = size\ of\ the\ compressed\ data \leq \mathcal{S}_{u}
\end{align*}
For data compression, we use the standard ZLIB library \cite{ZLIB:Deutsch}. ZLIB library ensures the least resource utilization during data compression.

\item \textbf{\textit{Feature 5 ($\Phi_5$)}:}  All data words from each SRAM chip are split into multiple blocks to extract this feature, where each block consists of $512$ consecutive data words. Then we compute the fractional value ($P_1$) of each block of data that exhibits logic ``1". We, then, calculate the standard deviation of $P_1$ calculated from each block. We select this standard deviation as feature $\Phi_5$. This feature captures the spatial locality of logic ``0" and logic ``1" of start-up data. A higher value of $\Phi_5$ signifies a larger spatial locality. Although we select the block size of 512, the manufacture may wish to select a different size that describes the best structural granularity in memory space. A smaller value of the block size might capture more spatial details; however, the $\Phi_5$ will also be largely influenced by the local noise if the block size is too small. We experimented with different block sizes and found that 512 provided the best result for memory classification. It is worth mentioning that $\Phi_3$ is similar to $\Phi_5$, where the block size of $\Phi_3$ is only one word. Hence, $\Phi_3$ captures finer grain spatial information more effectively. However, $\Phi_3$ may also capture the local noise information.

\item \textbf{\textit{Feature 6 ($\Phi_6$)}:} For each memory cell, we have collected SRAM data a total of 20 times and mark those memory cells as noisy if logic ``1" is observed 8 to 12 times. We marked those cells as noisy signature bits. For this feature, we counted the percentage of noisy signature bits. In a well-designed SRAM memory cell, the coupled inverters are highly matched, and corresponding signature bits are largely affected by the external/internal noises (e.g., voltage fluctuation, thermal noise, etc.). Furthermore, we believe that this feature can contribute highly to detect recycled memory chips. Over the usage, there will be more cells with large threshold voltage mismatch in recycled memory chips \cite{Recycled:Guin} and will produce large $PSNM_{noise}$ (see Sec. \ref{sec:distFactor}). Hence, a recycled SRAM chip should produce less noisy signature bits and reduce the value of $\Phi_6$ over time. 

\item \textbf{\textit{Feature 7 ($\Phi_7$)}:} 
In this feature, instead of accounting for the theoretical normal data distribution, we made a ($w_{l}+1$)-bin histogram. If a data-word ($\in D$) occupies a total $t$ bit of logic ``1", and then it is placed in $t$th histogram bin. The standard deviation of the bin size quantifies as the feature $\Phi_7$. If the distribution is normal (or Gaussian), then $\Phi_3$ and $\Phi_7$ should be approximately the same (also well-known as \textit{the normal approximation for probability histogram}). Hence, the $\Phi_7$ measures the skewness on word ($\in D$) distribution from the normal distribution.
\end{itemize}

We extract all these seven features from both fresh (i.e., new) and aged (i.e., recycled) SRAM chips. Then we show that these features form visually separated clusters in feature space depending on the SRAM module type (manufacturer ``A" vs. Manufacturer ``B", Part-number ``X" vs. Part-number ``Y", fresh vs. aged/recycled). 

In addition to above features, manufacturers may choose a different feature-set that describe their chips more concisely. Furthermore, the manufacturer may prefer a different set of data (e.g., error pattern by reducing latency parameters) to extract the more appropriate features \cite{DRAM_Host:Talukder}. However, when the manufacturer itself does not define the features and assign the responsibility to a third-party, one or few features might not obtain the exact electrical characteristics as intended due to the special modification at the architectural or layout level (which might not be known to the third-party). For example, bit-level scrambling in the data word may limit the usefulness of feature $\Phi_2$ \cite{TI:dataScrambling}. Nevertheless, this problem can be avoided by using conventional feature selection and dimension reduction techniques to select the most appropriate feature-set \cite{zhang:featSel, fernandez:featSel, Song:DimRed}. 

It is worth mentioning that the features described above only provide qualitative information of different physical properties of the SRAM chips; however, they do not provide any quantitative information. Furthermore, each feature described above might be impacted by combined information from multiple physical properties. For example, although $\Phi_5$ primarily varies from one memory class to another due to spatial variation, $\Phi_5$ might also be impacted by the address scrambling caused by the architectural difference in the address decoder \cite{Goor:AddressScramble}.

\subsection{Identifying Authentic Memory Chips}\label{sec:Framework}
Usually, memory chips with the same manufacturer and specification are labeled with a unique part-number; hence, to identify a memory authenticity, we need to identify the memory part-number. We propose a machine learning-based approach to classify the memory part-number after extracting features from the start-up data. However, the classification can be done with two different approaches- a) learning a binary classifier (positive vs. negative) for each class, and b) learning a one-class classifier for each class. In the first approach, we learn a binary classifier for each class to differentiate between positive samples and negative samples (i.e., authentic vs. counterfeit). This approach is only applicable when both positive and negative sample is available while training the classifier. Nonetheless, it is not a practical approach due to the enormous diversity in negative samples. Collecting negative samples from whole statistical distribution is not cost-effective and time-efficient. In the second approach, we do not need any samples from the negative class, and only positive samples are sufficient to learn the classifier. Recent studies show that \cite{DRAM_Host:Talukder,ML:Ahmadi, MemPAt:Rahman, ML:Sinanoglu, ML:Huang}, a one-class classifier is preferable for counterfeit IC detection as the statistical diversity of the counterfeit chips (negative class) is too large, and they can be introduced from a large number of sources (see Sec. \ref{sec:supplyChain}). Unfortunately, one-class classification is a complex statistical problem and might reduce the accuracy. Hence, we propose a two-step approach to solve this issue: 
\begin{enumerate}[leftmargin=*, topsep=0pt,itemsep=-1ex,partopsep=1ex,parsep=1ex]
  \item \textbf{Identifying manufacturer:} Different vendors use different memory cell designs, design flow, and possibly fabrication facilities. Furthermore, they may integrate different peripheral inside the memory; for example, altering row-decoder may alter apparent start-up data locality seen from outside of the memory. Hence, multiple sources may contribute to start-up data variation among SRAMs manufactured by different vendors. In other words, SRAMs for different manufacturers appeared to have a larger difference in their features (large inter-manufacturer feature distance), which ease identifying the SRAM manufacturer (e.g., manufactured by vendor ``A" or not). However, while training a binary-classifier, it is impossible to learn all the negative samples that the target vendor does not manufacture.  Therefore, we propose a one-class learner (e.g., one-class Neural Network, one-class SVM, SVDD, etc. \cite{DRAM_Host:Talukder,ML:Ahmadi, MemPAt:Rahman, ML:Sinanoglu, ML:Huang}) only to identify the manufacturer information. However, one may choose to train a binary-class classifier with all available negative samples along with a one-class classifier to improve the accuracy. 
  \item \textbf{Identifying part-number:} A manufacturer usually produces different memory chips with different specifications with different part-numbers. However, they may use the same design facility and similar peripherals for all of them, leading to a more subtle feature difference among memories. Fortunately, we can assume that a manufacturer can easily access all memories that they manufacture. Therefore, once the manufacturer is identified, the target manufacturer can easily provide a binary (target class vs. others) or a multi-class classifier to identify each memory part-number produced by them. As we mentioned earlier, the one-class classifier is a complex learning task; hence we should avoid it when we have access to the negative samples from the whole statistical distribution. In this particular scenario, one-class learning is more difficult as we have a smaller feature distance among part-numbers produced by the same manufacturer.
\end{enumerate}

In summary, for manufacturer identification, we recommend using a one-class classifier as it is difficult to collect samples from all manufacturers. However, for identifying part-number, we can safely assume that the manufacturer has access to samples from all part-number manufactured by itself. Therefore, we recommend using a binary (target class vs. others) or a multi-class classifier for identifying part-number. The one-class classifier is a complex statistical problem and requires large train samples. Unfortunately, we have only access to a limited number of samples from five major manufacturers. Therefore, for demonstration purposes, we avoid the one-class classifier for manufacturer identification and use a binary classifier for both manufacturer and part-number identification.

\subsection{Proposed Framework}\label{sec:propFramework}
\begin{figure}[ht!]
\centering
\includegraphics[width=0.75\textwidth, trim=0 6cm 0 0, clip]{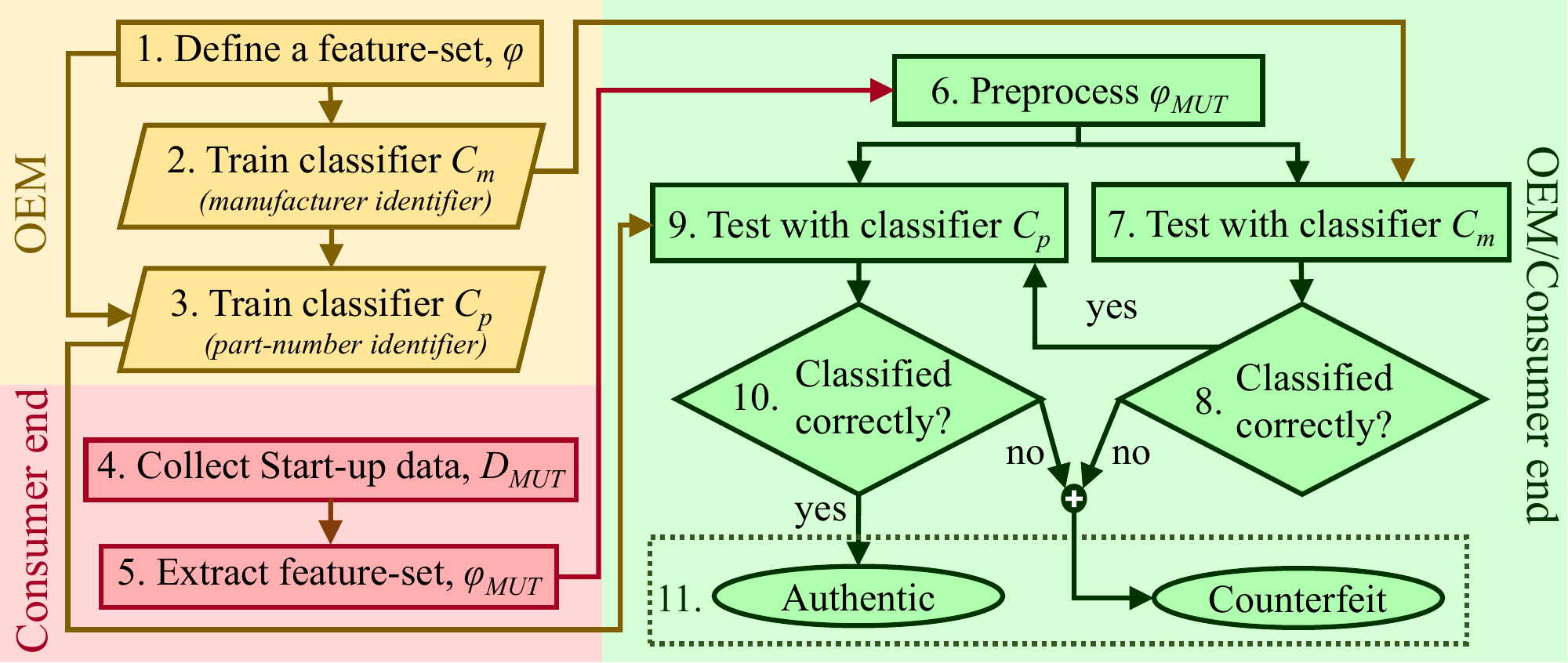} 
\caption{Proposed protocol to identify counterfeit SRAM.}
\label{fig:Protocol}
\end{figure}

We propose a machine learning-based algorithm that uses the device signature to verify the manufacturer and the part-number. Fig. \ref{fig:Protocol} represents the detailed framework of our proposed technique. Our proposed framework consists of eleven steps, and the order of the steps is numbered in Fig. \ref{fig:Protocol}. The steps provided in the yellow and red regions must be performed by the OEM and device owner (user/consumer), respectively. However, steps in the green region can be performed either by the user or by the manufacturer. Our proposed framework starts by defining a set of features (as explained in Sec. \ref{sec:FeatSelect}). Then, using a golden set\footnote{In the presence of an untrusted foundry, it is still possible to construct a golden sample set. An untrusted foundry can cause three types of possible counterfeiting- (\textbf{i}) introduce defective chips as fully functional, (\textbf{ii}) tamper the GDSII and introduce hardware trojan or security backdoor, and (\textbf{iii}) sell overproduced chips. While choosing the golden samples, it is easy to avoid defective chips if the design follows standard DFT (design for testing) and memory testing (such as memory built-in self-test or MBIST) techniques. On the other hand, tempering the GDSII requires R\&D effort to understand and modify the original design without altering the main chip functionality. Therefore, tampered chips can only be supplied to manufacturers (by the untrusted foundry) if the allowed GDSII to fabrication time is very long. Lastly, our proposed technique is not effective in identifying overproduced chips (see Sec. \ref{subsec:scope_lim}). As the feature-set extracted from the overproduced chips (with unmodified GDSII) should be the same as the authentic chip, there will be no impact on the classifier if the overproduced chips are used to train it.} of sample memory chips, the manufacturer needs to extract the feature set and train classifiers to identify counterfeit chips. The manufacturer can train the classifier in two steps: (\textbf{i}) learning manufacturer-specific property ($C_m$) and (\textbf{ii}) learning part-number-specific property ($C_p$). Manufacturing-specific property can be learned by a one-class classifier (i.e., only learning the target manufacturer) and might be assisted by a binary classifier (i.e., target manufacturer vs. others). For the second step, the manufacturer can train either a multi-class classifier for all part-numbers or a multiple binary (one vs. all) classifier for each part-number. By using publicly available information provided by the manufacturer, a user should be able to collect the signature from his sample and extract the feature-set. If the classifier information is available, the user can verify the chip authenticity by himself. Otherwise, the user can send the extracted feature-set to the manufacture, and the manufacturer can verify the authenticity of the test memory chip (verify on request). The start-up data collection process can vary depending on the evaluation platform and application; therefore, we leave the detailed implementation of the start-up data collection routine to the OEM/user. Note that our proposed protocol may be adjusted to meet the evaluation platform/manufacturer requirements.

\subsection{Identifying Recycled Memory Chips}\label{sec:RecMem}
Although identifying memory manufacturer and part-number can prevent many types of counterfeitings \cite{DRAM_Host:Talukder}, identifying memory manufacturer and part-number does not capture the recycled memory chips. Fortunately, the features we described in Sec. \ref{sec:FeatSelect} can also be used for identifying recycled memory chips. For example, the distribution of the 0's and 1's can be skewed over time due to the skewed distribution of 0's and 1's in functional memory usage, which can be easily captured by Feature 1 \cite{Recycled:Guin}. Additionally, we observe that the distribution of other features may help to identify recycled SRAM chips in extreme cases, i.e., when only the symmetric data patterns are used over functional memory usage (see Sec. \ref{sec:RecycleSRAM}).

\section{Result and Analysis} \label{sec:result}
In our experiment, we have collected SRAM start-up signatures to demonstrate our proposed technique \footnote{We have made our data publicly available at: \url{https://sourceforge.net/projects/authentimem/files/}}. Typically, the success of any machine learning (ML) model relies on the sample quality and sample size. However, it is difficult to collect data from a large set of sample chips in a lab environment and imitate all possible operating conditions. Therefore, we divide the data collection process into the following tasks:

\begin{enumerate}[leftmargin=*, topsep=0pt,itemsep=-1ex,partopsep=1ex,parsep=1ex]
    \item We used Arduino Due board \cite{arduino:Due} for collecting start-up data from SRAM chips. We have used 345 4-Mbit (256K$\times$16) SRAM chips from 5 different manufacturers and 23 different part-numbers (i.e., 23 memory classes). All of these 23 part-numbers are tabulated in Table \ref{Tab:chipList}. From now on, we will use the ``tag" (specified in Table \ref{Tab:chipList}) to recall a specific memory part-number/class. We have used 230 SRAM chips to train ML models (10 chips from each class) and 115 chips to test the model (5 chips from each class).
    \item We have collected data from both test chips and train chips at a nominal voltage (3.3V) and room temperature ($25^{\circ}$C). We used two different Arduino boards to emulate the platform variation among different embedded systems and utilized them to collect start-up signatures from test samples\footnote{We have not observed any visible impact from platform variation. Such observation is expected as the two testing platforms only differed by operating voltage (within 3.3$\pm$35mV), which is within the range of SRAM normal operating voltage  (3.0 to 3.6 mV). Most modern chips are equipped with a voltage clamp circuit and can control the internal voltage as long as the external voltage is within the range \cite{Maloney:clamp}.}. 
    \item We have used a one-vs-all binary classifier (positive vs. negative) for both manufacturer identification and part-number identification. As we explained in Sec. \ref{sec:Framework}, the one-classifier would be the best for the manufacturer identification. However, the one-class classification task is a complex statistical problem and might require a large number of samples to train the model.\label{task:3}
    \item Data noise can impact the classification models severely. To reduce noise, we collected start-up data from the same SRAM chips 20 times. We maintained a constant sampling interval of 2 minutes. We have shorted the power pin ($V_{CC}$) and other control pins of the SRAM chip with the ground within this time interval We maintained such settings using relay circuits (also controlled by the same Arduino Due board). This experimental setup should be sufficient to avoid the potential discharge inversion effect on the SRAM start-up state \cite{Liao:dischargeInversion}. We combined those 20 sets of data in a single set using the majority voting technique \cite{MajVoting}. 
    
    \begin{table}[ht!]
    \setcellgapes{2pt}
    \captionsetup{justification=centering, margin= 0cm}
    \caption{List of SRAM chips in experiment.} \label{Tab:chipList}
    \makegapedcells
    \centering
    \setlength\tabcolsep{3pt} 
    \resizebox{0.85\textwidth}{!}
    {
       \begin{tabular}{|c?c|c|c|c|c?c|c|c|c|c|c?c|c|c|c|c?c|c|c?c|c|c|c|} 
        \hline
        Manufacturer\footnotemark  & \multicolumn{5}{c?}{CY} & \multicolumn{6}{c?}{IDT} & \multicolumn{5}{c?}{ISSI} & \multicolumn{3}{c?}{AMI} & \multicolumn{4}{c|}{REA}                                                                       \\ 
        \hline
        {Part-number} &
        \rotatebox[origin=c]{90}{CY7C1041G30-10ZSXI} & \rotatebox[origin=c]{90}{CY7C1041CV33-20ZSXA} & \rotatebox[origin=c]{90}{CY7C1041G18-15ZSXI} & \rotatebox[origin=c]{90}{CY62147G30-55ZSXE} & \rotatebox[origin=c]{90}{CY62146EV30LL-45ZSXIT} & \rotatebox[origin=c]{90}{IDT71V416S10PHG8} &  \rotatebox[origin=c]{90}{IDT71V416S12PHG8} &  \rotatebox[origin=c]{90}{IDT71V416L15PHG8} &  \rotatebox[origin=c]{90}{IDT71V416S10PHGI} &   \rotatebox[origin=c]{90}{IDT71V416S12PHG} &    \rotatebox[origin=c]{90}{IDT71V416L15PHG} &  \rotatebox[origin=c]{90}{IS61LV25616AL-10TL} & \rotatebox[origin=c]{90}{IS61WV25616BLL-10TL} & \rotatebox[origin=c]{90}{IS61WV25616BLL-10TLI-TR} & \rotatebox[origin=c]{90}{IS61LV25616AL-10TLI} & \rotatebox[origin=c]{90}{IS61C25616AS-25TLI} & \rotatebox[origin=c]{90}{AS7C34098A-10TCN} &  \rotatebox[origin=c]{90}{AS7C34098A-10TIN} &    \rotatebox[origin=c]{90}{AS6C4016-55ZIN} &   \rotatebox[origin=c]{90}{RMLV0414EGSB-4S2\#AA1} & \rotatebox[origin=c]{90}{RMLV0414EGSB-4S2\#HA1} & \rotatebox[origin=c]{90}{RMLV0416EGSB-4S2\#AA1} & \rotatebox[origin=c]{90}{RMLV0416EGSB-4S2\#HA1}  \\ 
        \hline
        Tag                                                  & \rotatebox[origin=c]{90}{CY1} & \rotatebox[origin=c]{90}{CY2}  & \rotatebox[origin=c]{90}{CY3} & \rotatebox[origin=c]{90}{CY4}  & \rotatebox[origin=c]{90}{CY5} & \rotatebox[origin=c]{90}{IDT1} & \rotatebox[origin=c]{90}{IDT2} & \rotatebox[origin=c]{90}{IDT3} & \rotatebox[origin=c]{90}{IDT4} & \rotatebox[origin=c]{90}{IDT5} & \rotatebox[origin=c]{90}{IDT6} & \rotatebox[origin=c]{90}{ISSI1} & \rotatebox[origin=c]{90}{ISSI2} & \rotatebox[origin=c]{90}{ISSI3} & \rotatebox[origin=c]{90}{ISSI4} & \rotatebox[origin=c]{90}{ISSI5} & \rotatebox[origin=c]{90}{AMI1} & \rotatebox[origin=c]{90}{AMI2} & \rotatebox[origin=c]{90}{AMI3} & \rotatebox[origin=c]{90}{REA1} & \rotatebox[origin=c]{90}{REA2} & \rotatebox[origin=c]{90}{REA3} & \rotatebox[origin=c]{90}{REA4} \\
        \hline
        \end{tabular}
    }
    \end{table}
    \footnotetext{CY: Cypress Semiconductor; IDT: Integrated Device Technology; ISSI: Integrated Silicon Solution, Inc.; AMI: Alliance Memory, Inc.; REA: Renesas Electronics.}
    
    \item The variance error is expected when the sample size is too small \cite{Putten:bias-variance}. A model with high variance provides too much attention to the data that are trained with and prone to overfitting. Hence, to reduce the variance error in the trained model, we segmented the SRAM signature data in 16 chunks and virtually increased the sample count by treating each segment as an individual memory chip (i.e., extracting an individual set of features from each segment). However, in the inference phase, the class of a test sample is determined by the majority voting method using all 16 segments. If the same number of votes supports multiple class labels, the tie is broken by comparing the cumulative posterior probabilities\footnote{The posterior probability quantifies the confidence level of inferencing a sample to a particular class \cite{Hastie:ML}.} of all 16 segments. \label{task:5}
    \item To examine the temperature sensitivity of our proposed technique, we collected data from test samples at high temperatures ($\sim45^{\circ}$C) and validated the same trained model learned in task \ref{task:3}.
\end{enumerate}

\subsection{Visualizing Features} \label{sec:featVis}
\begin{figure*}[ht!]
\centering
\begin{subfigure}[]{0.49\textwidth}
    \includegraphics[width=1\textwidth, trim=1.7cm 0.3cm 1cm 0.7cm, clip]{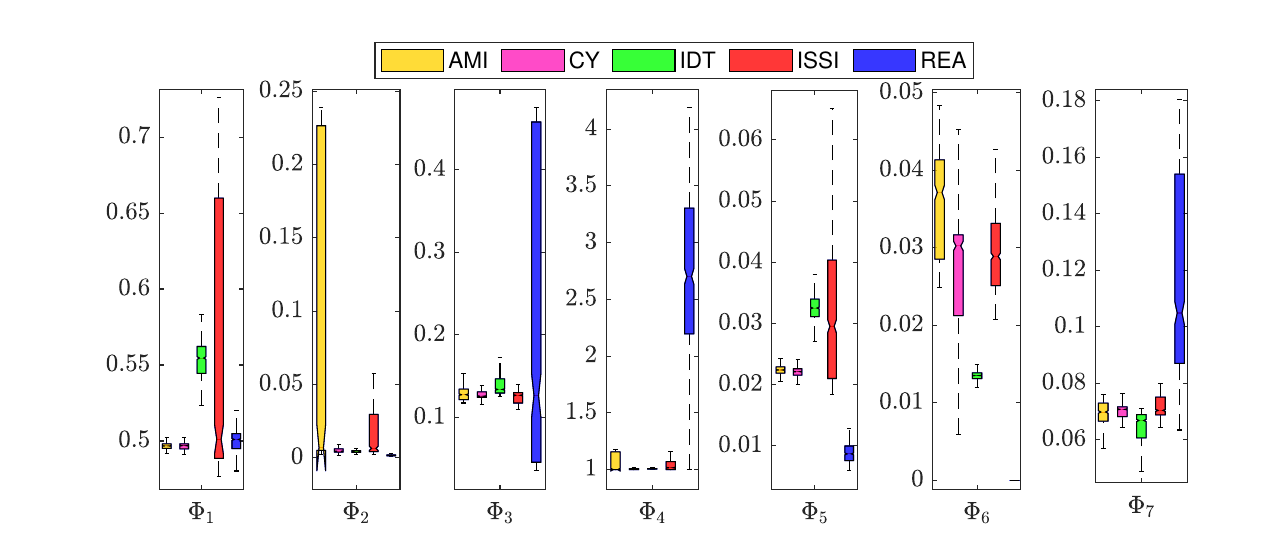} 
    \caption{By Manufacturer.}
    \label{fig:Feature_Distribution_by_vendor}
\end{subfigure}
\begin{subfigure}[]{0.49\textwidth}
    \includegraphics[width=1\textwidth, trim=1.6cm 0.3cm 1.2cm 0.7cm, clip]{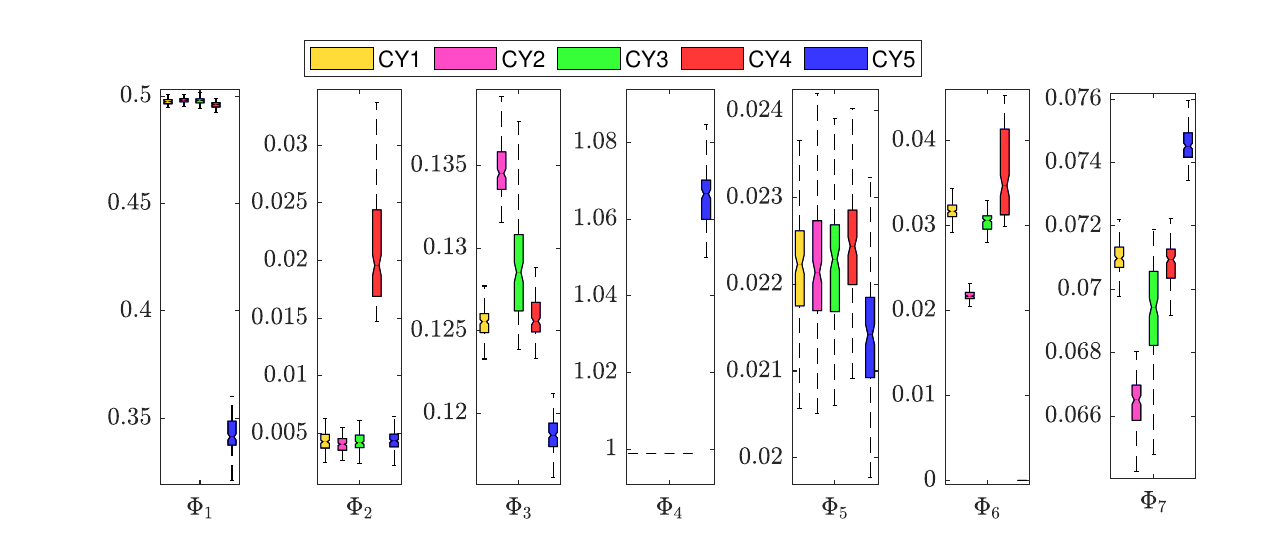}
    \caption{Cypress Semiconductor}
    \label{fig:Feature_Distribution_CY}
\end{subfigure}
\begin{subfigure}[]{0.49\textwidth}
    \includegraphics[width=1\textwidth, trim=1.7cm 0.3cm 0.7cm 0.5cm, clip]{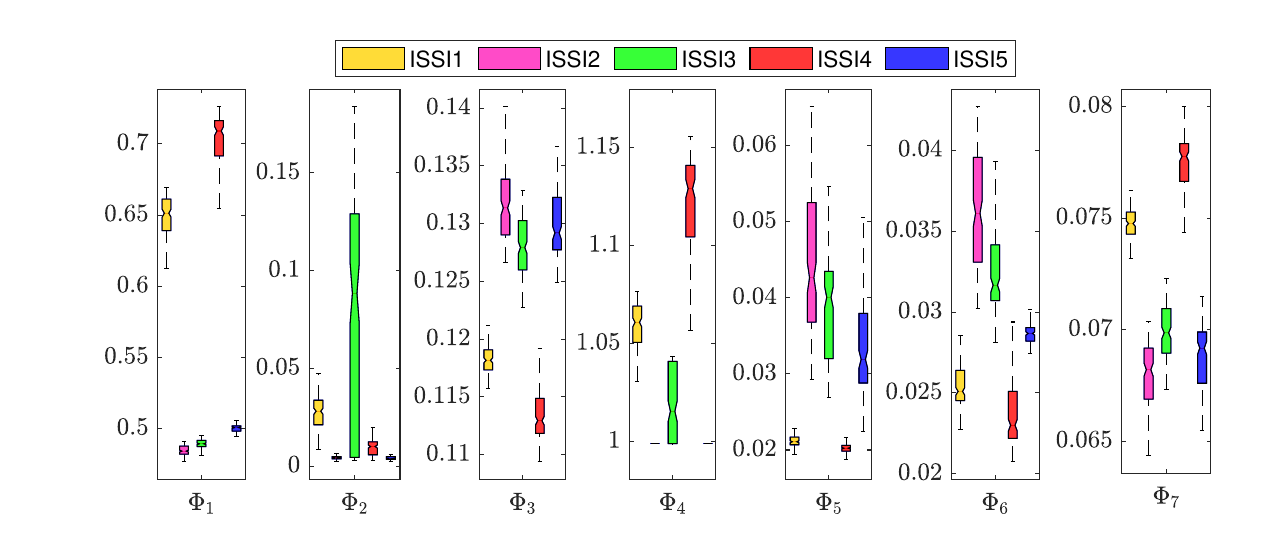}
    \caption{Integrated Silicon Solution, Inc.}
    \label{fig:Feature_Distribution_ISSI}
\end{subfigure}
\begin{subfigure}[]{0.49\textwidth}
    \includegraphics[width=1\textwidth, trim=1.5cm 0.3cm 0.9cm 0.5cm, clip]{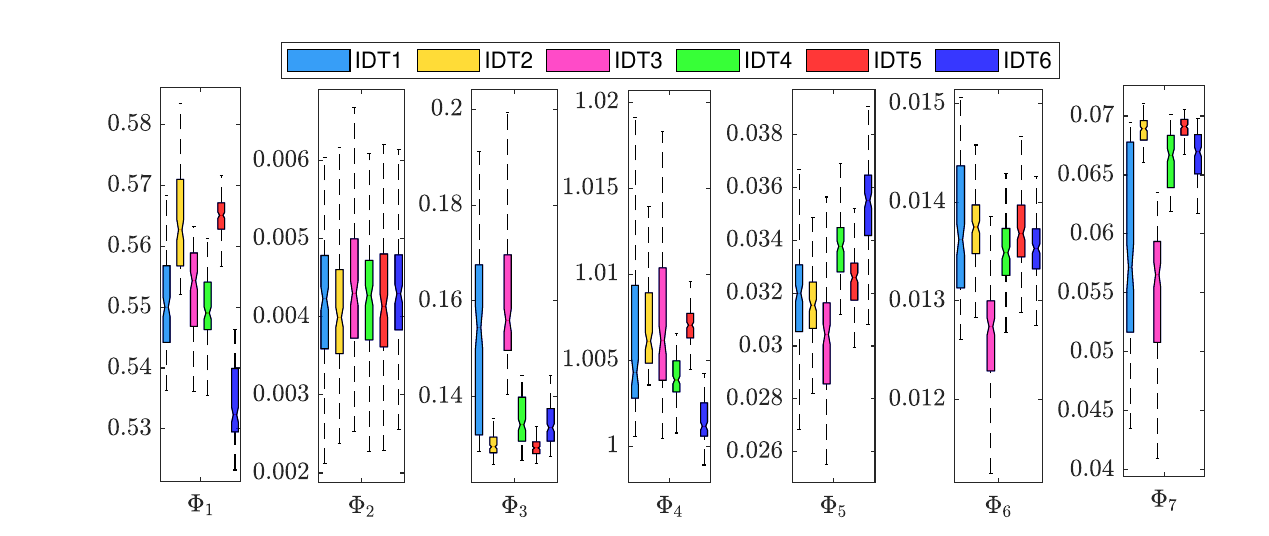}
    \caption{Integrated Device Technology}
    \label{fig:Feature_Distribution_IDT}
\end{subfigure}
\begin{subfigure}[]{0.49\textwidth}
    \includegraphics[width=1\textwidth, trim=1.7cm 0.3cm 0.7cm 0.5cm, clip]{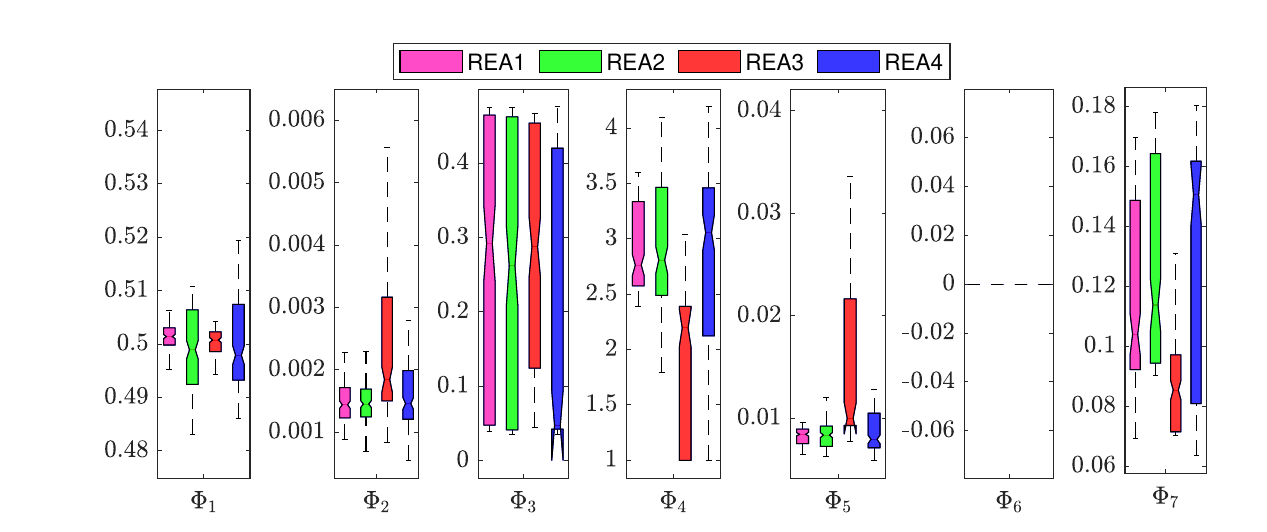}
    \caption{Renesas Electronics}
    \label{fig:Feature_Distribution_REA}
\end{subfigure}
\begin{subfigure}[]{0.49\textwidth}
    \includegraphics[width=1\textwidth, trim=1.3cm 0.3cm 0.9cm 0.5cm, clip]{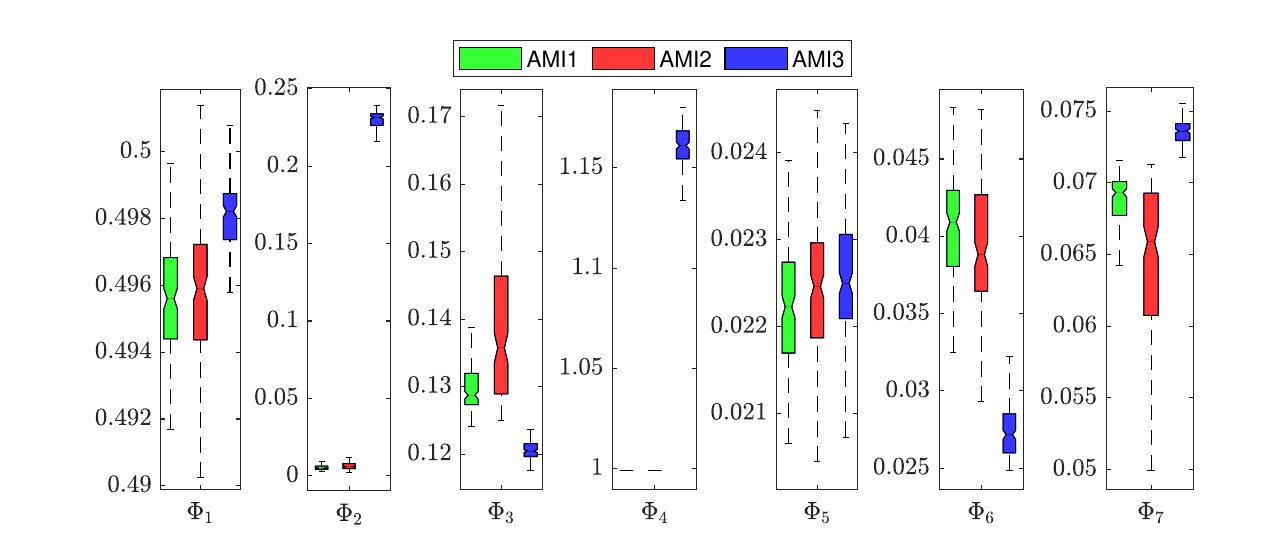}
    \caption{Alliance Memory}
    \label{fig:Feature_Distribution_AMI}
\end{subfigure}
\caption{Visualizing feature distribution by- (a) vendor, and (b)-(f) part-number.}
\label{Fig:Feature_Distribution}
\end{figure*}

The accuracy and efficiency of an ML algorithm largely depend on the quality of the features. Hence, to demonstrate the feature-merit (explained in Sec. \ref{sec:FeatSelect}), we have presented the feature distribution of train chip across different manufacturers and different part-numbers in Fig.\ref{Fig:Feature_Distribution}. The figure shows that most features are normally distributed (median is centered), and in many cases, at least one feature distribution of a particular class produces a clear visible separation with other classes (i.e., manufacturer ``A" vs. all and part-number ``X" vs. all). For example, in Fig. \ref{fig:Feature_Distribution_by_vendor}, the SRAM chips manufactured by Renesas Electronics are readily separable by the distribution of feature $\Phi_5$. Similarly, Fig. \ref{fig:Feature_Distribution_CY} demonstrates that SRAM chips from CY4 are easily distinguishable from the distribution of feature $\Phi_2$. Unfortunately, in our case, many of the classes can not be separated from other classes based on their feature distribution due to the complex interaction among those features. For example, feature $\Phi_1$ (number of 1's) and $\Phi_4$ (compression ratio) might have a close relation; for instance, if the signature data is highly random, the $\Phi_1$ should be close to 0.5, and $\Phi_4$ should be close to 1.

\begin{figure*}[ht!]
\centering
\begin{minipage}[]{.33\textwidth}
    \centering
    \captionsetup{justification=centering, margin=.0cm}
    \includegraphics[width=0.95\textwidth, trim=0.6cm 0.6cm 1.7cm 0.6cm, clip]{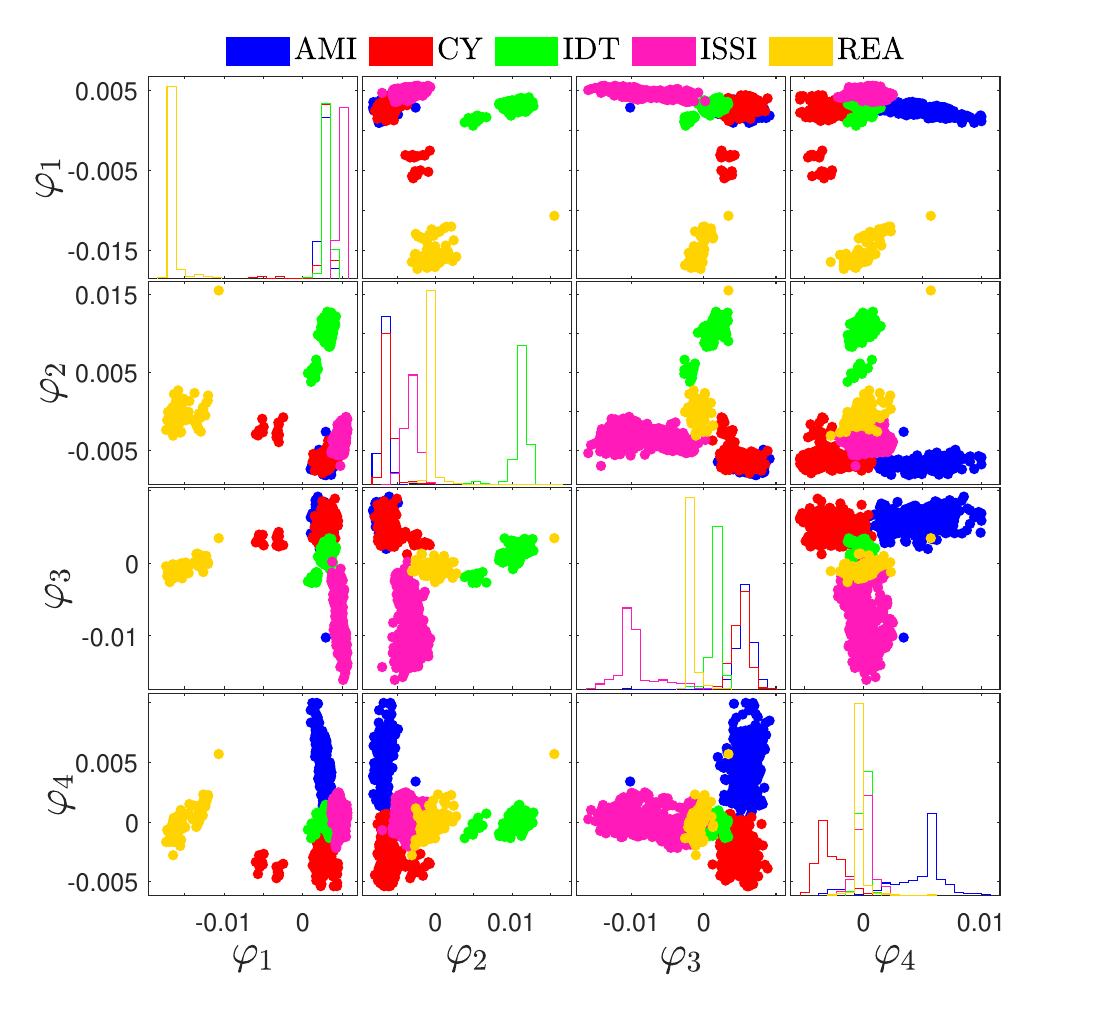}
    \subcaption{By Manufacturer}
    \label{fig:vendor_detection}
\end{minipage}%
\begin{minipage}[]{.33\textwidth}
    \centering
    \captionsetup{justification=centering, margin=0cm}
    \includegraphics[width=0.95\textwidth, trim=0.6cm 0.6cm 1.7cm 0.6cm, clip]{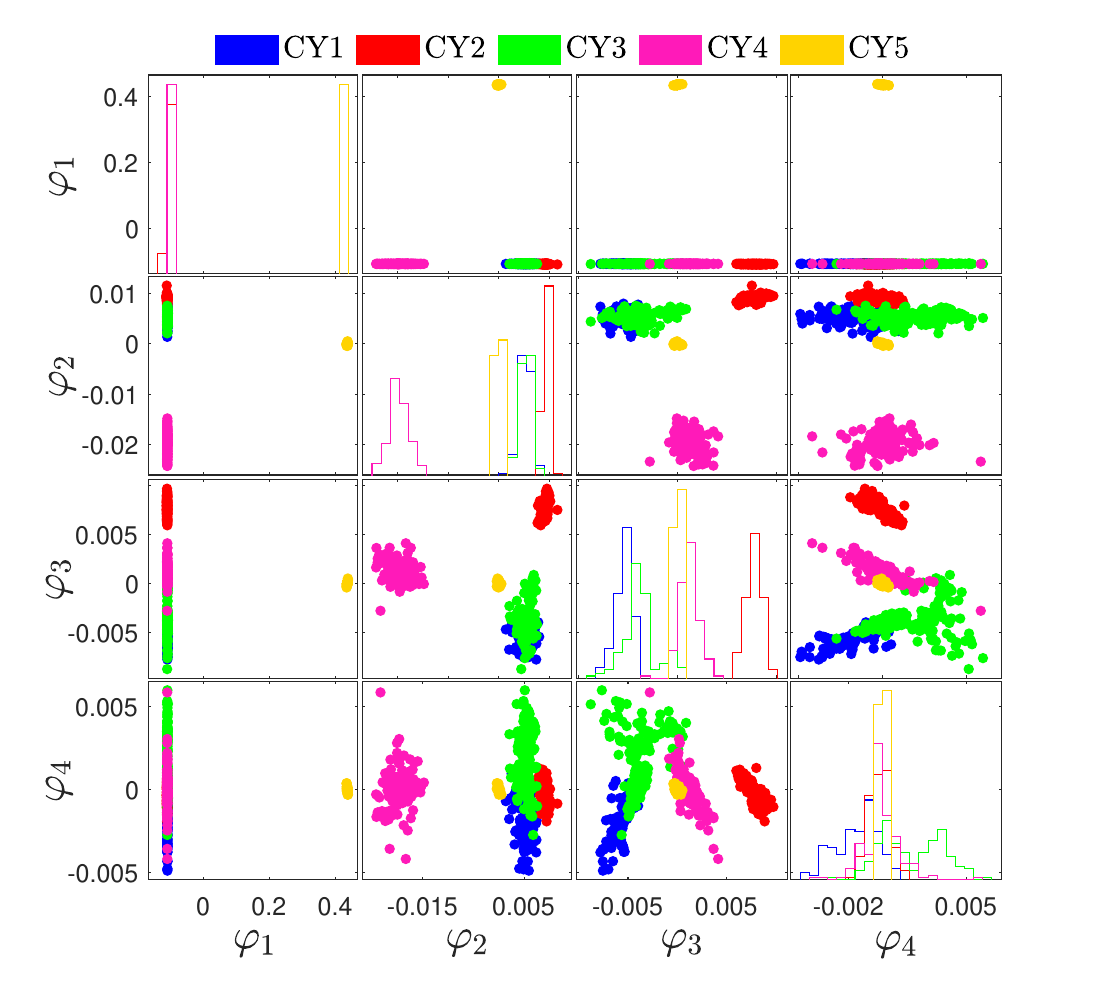}
    \subcaption{Cypress Semiconductor}
    \label{fig:spec_detection_CY}
\end{minipage}%
\begin{minipage}[]{.33\textwidth}
    \centering
    \captionsetup{justification=centering, margin=0cm}
    \includegraphics[width=0.99\textwidth, trim=0.3cm 0.5cm 1.6cm 0.5cm, clip]{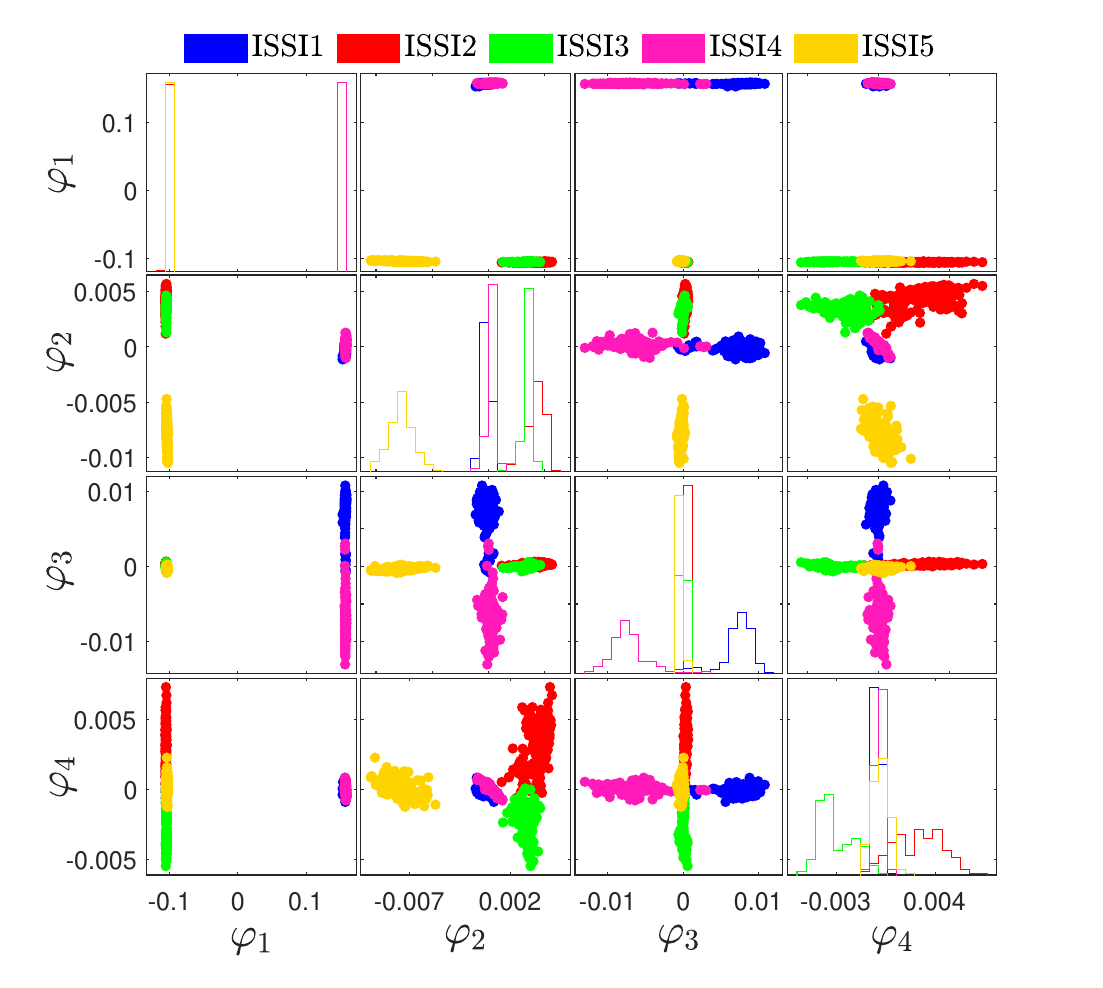}
    \subcaption{Integrated Silicon Solution, Inc.}
    \label{fig:spec_detection_ISSI}
\end{minipage}%
\allowbreak%
\vspace{5pt}
\begin{minipage}[]{.25\textwidth}
    \centering
    \captionsetup{justification=centering, margin=0cm}
    \includegraphics[width=0.97\textwidth, trim=0cm 0cm 1.2cm 0.1cm, clip]{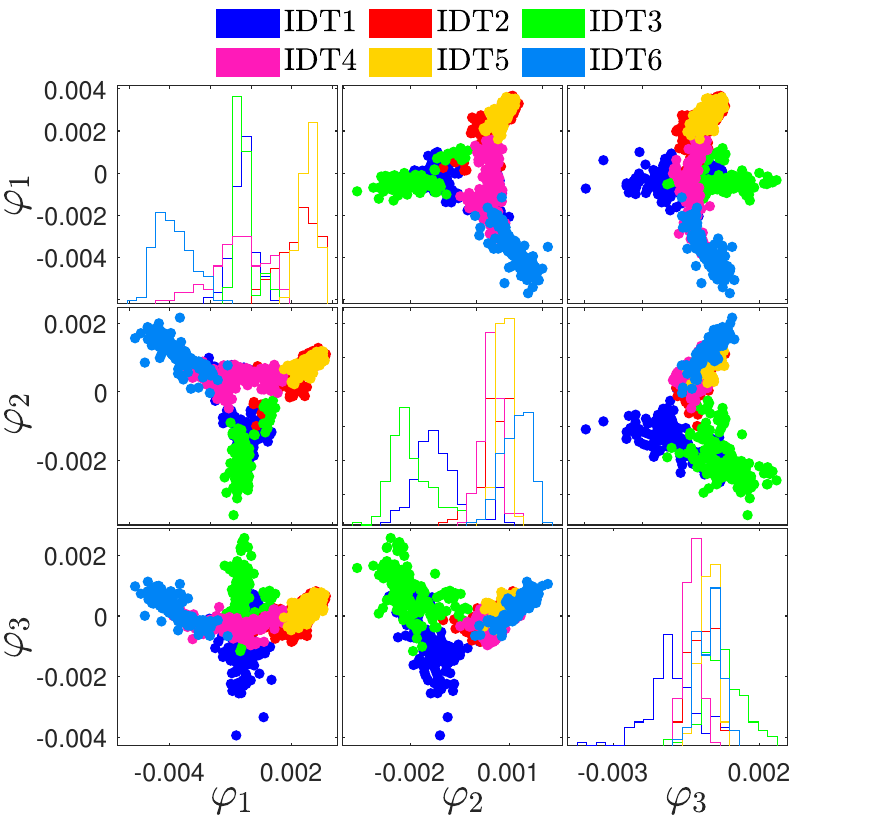} 
    \subcaption{Integrated Device Technology}
    \label{fig:spec_detection_IDT}
\end{minipage}%
\hspace{0.2cm}
\begin{minipage}[]{.25\textwidth}
    \centering
    \captionsetup{justification=centering, margin=0cm}
    \includegraphics[width=0.97\textwidth, trim=0cm 0.1cm 1.3cm -0.5cm, clip]{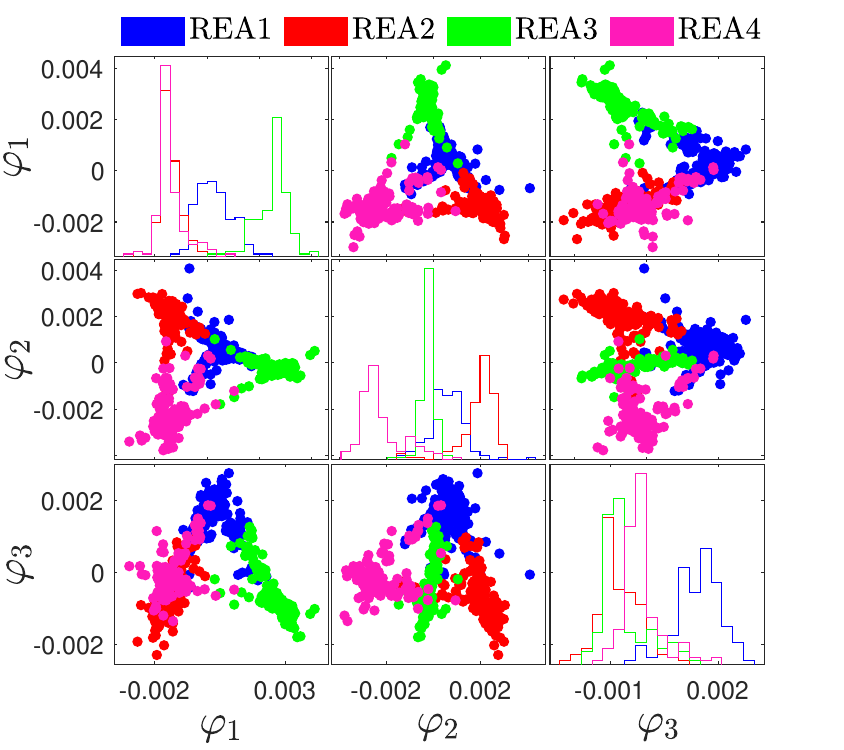} 
    \subcaption{Renesas Electronics}
    \label{fig:spec_detection_REA}
\end{minipage}%
\hspace{0.22cm}
\begin{minipage}[]{.10\textwidth}
    \centering
    \captionsetup{justification=centering, margin=0cm}
    \includegraphics[width=0.98\textwidth, trim=0.4cm -2.5cm 9.1cm 1cm, clip]{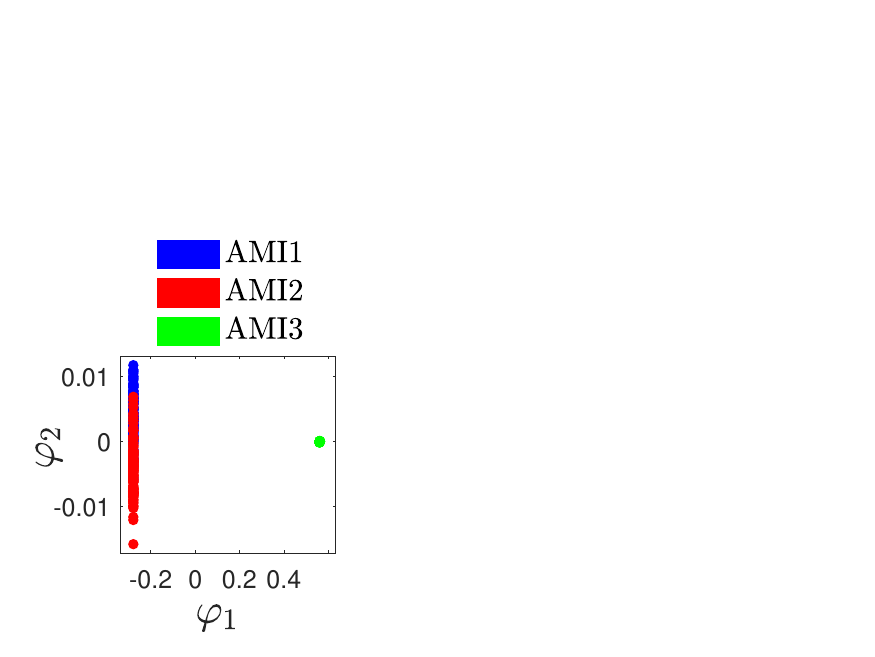} 
    \subcaption{Alliance \\Memory}
    \label{fig:spec_detection_AMI}
\end{minipage}%
\caption{Representation of SRAM memory in feature-space, clustered by- (a) vendor, and (b)-(f) part-number.}
\label{fig:vend_spec_detection}
\end{figure*}

For such cases, the class separability can still be visualized if the current feature-space ($\Phi$-space) is transferred to a new feature-space ($\varphi$-space), where the $\varphi_i = f(\Phi_1, \Phi_2, \cdots, \Phi_n)$. 
In our experiment, we have used generalized discriminant analysis (GDA) \cite{Baudat:GDA} to transform the $\Phi$-space to $\varphi$-space, where data points are linearly separable at $\varphi$-space. GDA\footnote{The reference implementation of GDA is available at: \url{https://github.com/mhaghighat/gda}} is a supervised machine learning technique to find a reduced set of features that preserves the maximum separability among the classes. This reduced set of features is related to the old feature space by a non-linear kernel function. In our experiment, we have used an \textit{RBF} kernel function \cite{DRAM_Host:Talukder}. The \textit{RBF} functions' parameter ($\gamma$) is determined by the 10-fold cross-validation method and ensured minimum Euclidean distance between samples and corresponding centroids \cite{ranganathan_cross-validation_2019}. Fig. \ref{fig:vend_spec_detection} represents the test memory chips in $\varphi$-space (in 2D projection) and demonstrates the manufacturer and part-number separability. Each dot in Fig. \ref{fig:vend_spec_detection} represents each memory segment as explained in task-\ref{task:5}. Those two figures demonstrate that memory classes (manufacturer ``A" vs. ``B" and part-number ``X" vs. ``Y") are fairly distinguishable in at least one 2D projection of the $\varphi$-space. While transforming the feature-space of a $K$-class problem, it is worth mentioning that at most $K-1$ dimensions are required in the new feature-space without losing any information of class separability \cite{Hastie:ML}. However, for IDT, adding more than three dimensions (Fig. \ref{fig:spec_detection_IDT}) only adds very small details on class separability (which is not recognizable from visual appearance). However, we have used $K-1$ dimensional new space for a $K$-class problem for other cases in Fig. \ref{fig:vend_spec_detection}.

Note that some overlapping between multiple classes is still visible in the $\varphi$-space due to the random process variation. However, such overlapping can be reduced by further optimizing the RBF parameters, given that more train samples are available (we have only ten samples from each part-number). While classifying the test memory chips, the impact of such overlap is minimized by assigning equal weight on all 16 segments of the chip and casting a ``vote" from each segment.
\subsection{Labeling Test Memory Chips} \label{sec:testchipLab}
Although the GDA can be used for both visualization and classification tasks, GDA is not ideal for a small sample size. Fortunately, the ensemble learning technique can still perform reasonably better even with a small set of samples \cite{Dietterichl:Ensemble}. In the ensemble technique, multiple base models are learned with different configurations, and then the output label of the test sample is determined based on the vote cast by each model. Although several ensemble algorithms are available, we have used the bagging (\textbf{b}ootstrap \textbf{agg}regat\textbf{ing}) method in our experiment. The bagging method is similar to other ensemble methods, except the base model is trained with a different set of train data (sampled with replacement). The bagging method has the inherent ability to reduce the variance error of the trained model and can out-perform other ML algorithms when the train sample size is small \cite{bagging:khan}. The detailed construction of the algorithm is out of the scope of this paper.

In our experiment, we have trained multiple ensemble models using different base classifiers (e.g., SVM, Decision Tree, Naive Bayes, Discriminant Analysis, Kernel, etc.), and the best model is chosen based on the 10-fold cross-validation score. Then we generated the test score based on our test samples. We represented the test score in Table \ref{Tab:AccM} and \ref{Tab:AccP}. The table presents four types of test scores: \textit{Precision} ($P$), \textit{Recall} ($R$), $F_{1}$ score, and accuracy, which are defined by Eq. \ref{eqn:Precision}, \ref{eqn:Recall}, \ref{eqn:f1}, and \ref{eqn:Acc}, respectively. $P$ quantifies the trained model's accuracy out of all predicted positives, and the $R$ computes the fraction of positives that the model captures correctly. On the other hand, $F_{1}$ score is the harmonic mean of the $P$ and $R$. For an ideal case, all of these test scores should be close to 1. Note that, the accuracy is not a very useful metric when the test samples from the positive and negative classes are not equal (unbalanced data). In our experimental setup, the number of test samples for binary (one vs. all) classifiers is unbalanced; hence, we emphasize the $P$, $R$, and $F_{1}$ score in our discussion.

\begin{equation} \label{eqn:Precision}
Precision\ (P) = \frac{tp }{tp + fp} \\
\end{equation}
\begin{equation} \label{eqn:Recall}
Recall\ (R) = \frac{tp}{tp+fn}
\end{equation}
\begin{equation} \label{eqn:f1}
F_1 = \frac{2}{(P)^{-1}+(R)^{-1}} 
\end{equation}
\begin{equation} \label{eqn:Acc}
A = \frac{tp + tn}{tp + tn + fp +fn}
\end{equation}
\begin{minipage}{50em}
Where,\\
\hspace*{28pt} $tp$ = True positive; \hspace*{12pt} $tn$ = True negative; \hspace*{12pt} $fp$ = False positive; \hspace*{12pt} $fn$ = False negative 
\vspace{8pt}
\end{minipage}

We have trained our binary model by utilizing the samples from the target class and the samples from the outlier class (i.e., not belong to the target class). Target class implies manufacturer (or part-number), which is targeted to separate from other manufacturers (or part-numbers). Note that we can either consider the target class as the positive class or the outlier class as the positive class in Eq. \ref{eqn:Precision}, \ref{eqn:Recall}, \ref{eqn:f1}, and \ref{eqn:Acc}. Depending on the definition of positive class, the $P$, $R$, and $F_{1}$ score can be different for unbalanced test samples. We focus on the test scores produced by considering the target class as the positive class as it delivers the worse set of test scores.

Table \ref{Tab:AccM} and \ref{Tab:AccP} presents a single accuracy score and two sets of $P$, $R$, and $F_{1}$ score for manufacturer and part-number identification considering both objectives as discussed above. In Table \ref{Tab:AccM}, the $1^{st}$ row represents the target manufacturer, and the $2^{nd}$ row represents the corresponding accuracy score. Row 3, 4, and 5 represent the $P$, $R$, and $F_{1}$ score considering the target class as the positive class. Similarly, row 6, 7, and 8 represent the $P$, $R$, and $F_{1}$ score considering the outlier class as the positive class. Column 2--6 represents the classifier score for each manufacturer, and column 7 ($\mu^{V}$) represents the average classification score considering all manufacturers. The table shows that the average test scores are $\geq0.92$\% (positive class = target class), which is promising considering such a small number of samples. However, the classification score is a little lower for CY and AMI than the other manufacturers, resulting from the fact that CY and AMI slightly overlap in feature space (blue and red dots in \ref{fig:vendor_detection}). However, the classification scores can be improved by adding more samples and further optimization of the classifiers.

\begin{table}[ht!]
\setcellgapes{2pt}
\captionsetup{justification=centering, margin= 0cm}
\caption{Classification scores (at nominal temperature) for SRAM manufacturer identification.} \label{Tab:AccM}
\makegapedcells
\centering
\setlength\tabcolsep{3pt} 
\resizebox{0.5\textwidth}{!}
{
    \begin{tabular}{|c?c?c|c|c|c|c|c|} 
    \hline
    \multicolumn{2}{|c?}{Manufacturer} & CY   & IDT  & ISSI & AMI  & REA & $\mu^V$  \\ 
    \thickhline
    \multicolumn{2}{|c?}{$A$}     & 0.93 & 1.00 & 0.99 & 0.97 & 0.99 & 0.98   \\ 
    \thickhline
    \multirow{3}{*}{Target Class} & $P$       & 0.87 & 1.00 & 1.00 & 0.87 & 1.00 & 0.95   \\ 
    \cline{2-8}
                       & $R$       & 0.80 & 1.00 & 0.96 & 0.87 & 0.95 & 0.92   \\ 
    \cline{2-8}
                       & $F_1$      & 0.83 & 1.00 & 0.98 & 0.87 & 0.97 & 0.93   \\ 
    \thickhline
    \multirow{3}{*}{Outlier} & $P$       & 0.98 & 1.00 & 0.99 & 0.98 & 0.99 & 0.98   \\ 
    \cline{2-8}
                       & $R$       & 0.98 & 1.00 & 1.00 & 0.98 & 1.00 & 0.99   \\ 
    \cline{2-8}
                       & $F_1$      & 0.98 & 1.00 & 0.99 & 0.98 & 0.99 & 0.99   \\
    \hline
    \end{tabular}
}
\end{table}

\begin{table}[ht!]
\setcellgapes{2pt}
\captionsetup{justification=centering, margin= 0cm}
\caption{Classification scores (at nominal temperature) for SRAM part-number identification.} \label{Tab:AccP}
\makegapedcells
\centering
\setlength\tabcolsep{2pt} 
\resizebox{0.98\textwidth}{!}
{
    \begin{tabular}{|c?c?c|c|c|c|c|c?c|c|c|c|c|c|c?c|c|c|c|c|c?c|c|c|c?c|c|c|c|c?c|} 
    \hline
    \multicolumn{2}{|c?}{Manufacturer} & \multicolumn{6}{c?}{CY}                 & \multicolumn{7}{c?}{IDT}                       & \multicolumn{6}{c?}{ISSI}               & \multicolumn{4}{c?}{AMI}  & \multicolumn{5}{c?}{REA}         & \rtt{$\mu^{V}$}    \\ 
    \thickhline
    \multicolumn{2}{|c?}{Tag}    & \rtt{CY1}    & \rtt{CY2}    & \rtt{CY3}    & \rtt{CY4}    & \rtt{CY5}    & \rtt{$\mu^{M}$}   & \rtt{IDT1}    & \rtt{IDT2}    & \rtt{IDT3}    & \rtt{IDT4}    & \rtt{IDT5}    & \rtt{IDT6}    & \rtt{$\mu^{M}$}   & \rtt{ISSI1}    & \rtt{ISSI2}    & \rtt{ISSI3}    & \rtt{ISSI4}    & \rtt{ISSI5}    & \rtt{$\mu^{M}$}   & \rtt{AMI1}    & \rtt{AMI2}    & \rtt{AMI3}    & \rtt{$\mu^{M}$}   & \rtt{REA1}    & \rtt{REA2}    & \rtt{REA3}    & \rtt{REA4}    & \rtt{$\mu^{M}$}   & \rtt{$-$}     \\ 
    \thickhline
    \multicolumn{2}{|c?}{$A$}  &   \rtt{0.88} & \rtt{1.00} & \rtt{0.88} & \rtt{1.00} & \rtt{1.00} & \rtt{0.95} & \rtt{0.87} & \rtt{0.87} & \rtt{1.00} & \rtt{0.77} & \rtt{0.90} & \rtt{0.77} & \rtt{0.86} & \rtt{1.00} & \rtt{0.92} & \rtt{0.92} & \rtt{1.00} & \rtt{1.00} & \rtt{0.97} & \rtt{0.73} & \rtt{0.73} & \rtt{1.00} & \rtt{0.82} & \rtt{0.65} & \rtt{0.65} & \rtt{0.85} & \rtt{0.90} & \rtt{0.76} & \rtt{0.88}  \\ 
    \thickhline
    \multirow{3}{*}{\rtt{Target Class}} & $P$       & \rtt{0.75} & \rtt{1.00} & \rtt{0.67} & \rtt{1.00} & \rtt{1.00} & \rtt{0.88} & \rtt{1.00} & \rtt{1.00} & \rtt{1.00} & \rtt{0.33} & \rtt{1.00} & \rtt{\tcb{0.42}} & \rtt{0.79} & \rtt{1.00} & \rtt{0.80} & \rtt{0.80} & \rtt{1.00} & \rtt{1.00} & \rtt{0.92} & \rtt{\tcb{0.56}} & \rtt{1.00} & \rtt{1.00} & \rtt{0.85} & \rtt{\tcb{0.33}} & \rtt{\tcb{0.33}} & \rtt{1.00} & \rtt{0.71} & \rtt{0.60} & \rtt{0.81}  \\ 
    \cline{2-31}
                       & $R$       & \rtt{0.60} & \rtt{1.00} & \rtt{0.80} & \rtt{1.00} & \rtt{1.00} & \rtt{0.88} & \rtt{\tcb{0.20}} & \rtt{\tcb{0.20}} & \rtt{1.00} & \rtt{\tcb{0.40}} & \rtt{\tcb{0.40}} & \rtt{1.00} & \rtt{0.53} & \rtt{1.00} & \rtt{0.80} & \rtt{0.80} & \rtt{1.00} & \rtt{1.00} & \rtt{0.92} & \rtt{1.00} & \rtt{\tcb{0.20}} & \rtt{1.00} & \rtt{0.73} & \rtt{\tcb{0.40}} & \rtt{\tcb{0.40}} & \rtt{\tcb{0.40}} & \rtt{1.00} & \rtt{0.55} & \rtt{0.72}  \\ 
    \cline{2-31}
                       & $F_1$      & \rtt{0.67} & \rtt{1.00} & \rtt{0.73} & \rtt{1.00} & \rtt{1.00} & \rtt{0.88} & \rtt{0.33} & \rtt{0.33} & \rtt{1.00} & \rtt{0.36} & \rtt{0.57} & \rtt{0.59} & \rtt{\textcolor{cyan}{\textbf{0.53}}} & \rtt{1.00} & \rtt{0.80} & \rtt{0.80} & \rtt{1.00} & \rtt{1.00} & \rtt{0.92} & \rtt{0.71} & \rtt{0.33} & \rtt{1.00} & \rtt{0.68} & \rtt{0.36} & \rtt{0.36} & \rtt{0.57} & \rtt{0.83} & \rtt{0.53} & \rtt{0.71}  \\ 
    \thickhline
    \multirow{3}{*}{\rtt{Outlier}} & $P$       & \rtt{0.90} & \rtt{1.00} & \rtt{0.95} & \rtt{1.00} & \rtt{1.00} & \rtt{0.97} & \rtt{0.86} & \rtt{0.86} & \rtt{1.00} & \rtt{0.88} & \rtt{0.89} & \rtt{1.00} & \rtt{0.92} & \rtt{1.00} & \rtt{0.95} & \rtt{0.95} & \rtt{1.00} & \rtt{1.00} & \rtt{0.98} & \rtt{1.00} & \rtt{0.71} & \rtt{1.00} & \rtt{0.90} & \rtt{0.79} & \rtt{0.79} & \rtt{0.83} & \rtt{1.00} & \rtt{0.85} & \rtt{0.93}  \\ 
    \cline{2-31}
                       & $R$       & \rtt{0.95} & \rtt{1.00} & \rtt{0.90} & \rtt{1.00} & \rtt{1.00} & \rtt{0.97} & \rtt{1.00} & \rtt{1.00} & \rtt{1.00} & \rtt{0.84} & \rtt{1.00} & \rtt{0.72} & \rtt{0.93} & \rtt{1.00} & \rtt{0.95} & \rtt{0.95} & \rtt{1.00} & \rtt{1.00} & \rtt{0.98} & \rtt{0.60} & \rtt{1.00} & \rtt{1.00} & \rtt{0.87} & \rtt{0.73} & \rtt{0.73} & \rtt{1.00} & \rtt{0.87} & \rtt{0.83} & \rtt{0.92}  \\ 
    \cline{2-31}
                       & $F_1$      & \rtt{0.95} & \rtt{1.00} & \rtt{0.92} & \rtt{1.00} & \rtt{1.00} & \rtt{0.97} & \rtt{0.93} & \rtt{0.93} & \rtt{1.00} & \rtt{0.86} & \rtt{0.94} & \rtt{0.84} & \rtt{0.91} & \rtt{1.00} & \rtt{0.95} & \rtt{0.95} & \rtt{1.00} & \rtt{1.00} & \rtt{0.98} & \rtt{0.75} & \rtt{0.83} & \rtt{1.00} & \rtt{0.86} & \rtt{0.76} & \rtt{0.76} & \rtt{0.91} & \rtt{0.93} & \rtt{0.84} & \rtt{0.92}  \\
    \hline
    \end{tabular}
}
\end{table}

In Table \ref{Tab:AccP}, we have presented the classification score for the part number identification, where the $2^{nd}$ row represents the target part-number. Note that, $\mu^{M}$ represents the average classification score over the corresponding manufacturer, and the $\mu^{V}$ columns represents the average classification score over all manufacturers. Similar to Table \ref{Tab:AccM}, rows 4--9 of table \ref{Tab:AccP} represent two sets of $P$, $R$, and $F_{1}$ scores. Unlike manufacturer identification, the part-number classification score for some manufactures is not up to the mark; especially, the $P$ or the $R$ (and corresponding $F_{1}$ score) scores to identify a few part-numbers of IDT, REA, and AMI are unacceptably low (shown in red). Nevertheless, such low test scores can be explained from multiple perspectives. For example, the model used to classify manufacturers trained based on 40--60 samples per class; however, due to the extremely limited number of samples from each part-number (10 from each), it is harder to learn part-number classifiers. Besides, the differences among a few memory part-numbers, especially from IDT, REA, and AMI, are not well-understood from their electrical characteristics mentioned in the datasheets. For example, the only noticeable difference between IDT2 and IDT5 is how they are packed during shipping (tube/tray vs. tape/reel). Hence, these two part-numbers might be equivalent based on their electrical characteristics. Similarly, the following pair of the part-numbers- (IDT3, IDT6), (REA1, REA2), and (REA3, REA4) do not have any recognisable difference other than their packing method. Hence, to extract the perfect set of features to differentiate those chips (IDT2 vs. IDT5, IDT3 vs. IDT6, REA1 vs. REA2, and REA3 vs. REA4), we might require more detailed information about the chip characteristics. On the other hand, the IDT1 and IDT4 memory chips are only differed by the temperature grade, and possibly have only difference in their die packaging along with some minor fabrication imperfections \cite{mishra:tempGrade}. Hence IDT1 and IDT4 may have very subtle differences due to the possible similarity in die architectural, layout, and systematic process variation. We found the similar problem for AMI1 and AMI2, which are also only differed by the temperature grade \footnote{IDT1 and AMI1 are commercial grade (supports $0^{\circ}$ to $+70^{\circ}$C operating temperature); whereas, the  IDT4 and AMI2 are industrial grade (supports $-40^{\circ}$ to $+85^{\circ}$C operating temperature).}. Note that, the difference between IDT1 and IDT4 (or, between AMI1 and AMI2) might still be captured by using more train samples. Additionally, if we have more detailed information on chip design, we may be able to identify the subtle difference due to die packaging. For example, the die packaging and wire bonding should impact the characteristics of chips IOs; therefore, the peripheral circuitry of memory chips communicating with IOs should have more impact due to the difference in die packaging. Hence, a feature that captures the memory peripheral characteristics should be more suitable to capture the subtle variation due to die package variation. Unfortunately, understanding the memory peripheral characteristics requires detailed design information on peripheral design, which is only available to memory manufacturers.

In Table \ref{Tab:AccHT}, we have also presented the summary result (only average test score) by changing the operating temperature of the test samples to $\sim45^{\circ}$C. The $2^{nd}$ and last row of Table \ref{Tab:AccHT} represents the average score for the manufacturer and part-number detection (respectively) from all manufacturers. On the other hand, rows 3--7 represent the average score for part-number detection from the corresponding manufacturer and the row 8 represents the average score for part-number detection over all manufacturers. From Table \ref{Tab:AccM}, \ref{Tab:AccP} and \ref{Tab:AccHT}, it is apparent that our proposed technique is not very sensitive to temperature. The temperature insensitivity of our selected features is reasonable; previous work shows that varying $+60^{\circ}$C only changes the SRAM start-up data by $\sim12$\% \cite{Premalatha:SRAMTemp}.

\begin{table}[ht!]
\setcellgapes{2pt}
\captionsetup{justification=centering, margin= 0.5cm}
\caption{Arithmetic mean of classification scores (at high temperature)} \label{Tab:AccHT}
\makegapedcells
\centering
\setlength\tabcolsep{3pt} 
\resizebox{0.55\textwidth}{!}
{
   \begin{tabular}{|c|c?c?c|c|c?c|c|c|} 
    \hline
    \multicolumn{2}{|c?}{\multirow{2}{*}{\begin{tabular}[c]{@{}c@{}}Classification\\goal\end{tabular}}} & \multirow{2}{*}{$A$} & \multicolumn{3}{c?}{Target Class} & \multicolumn{3}{c|}{Outlier}  \\ 
    \Cline{1pt}{4-9}
    \multicolumn{2}{|c?}{}                  &                    & $P$    & $R$    & $F_1$        & $P$    & $R$    & $F_1$        \\ 
    \thickhline
    \multicolumn{2}{|c?}{Manufacturer ($\mu^V$)}               & 0.97               & 0.93 & 0.94 & 0.93      & 0.99 & 0.98 & 0.98      \\ 
    \thickhline
    \multirow{6}{*}{\rtt{\begin{tabular}[c]{@{}c@{}}Part-number\end{tabular}}}& CY ($\mu^M$)               & 0.97               & 0.93 & 0.96 & 0.93      & 0.99 & 0.97 & 0.98      \\ 
    \cline{2-9}
                         & IDT ($\mu^M$)              & 0.81               & \tcb{0.54} & \tcb{0.47} & \tcb{0.43}      & 0.90 & 0.88 & 0.88      \\ 
    \cline{2-9}
                         & ISSI ($\mu^M$)             & 0.95               & 0.93 & 0.88 & 0.87      & 0.97 & 0.97 & 0.97      \\ 
    \cline{2-9}
                         & AMI ($\mu^M$)              & 0.82               & 0.76 & 0.73 & 0.72      & 0.89 & 0.87 & 0.86      \\ 
    \cline{2-9}
                         & REA ($\mu^M$)              & 0.73               & \tcb{0.58} & \tcb{0.55} & \tcb{0.48}      & 0.85 & 0.78 & 0.81      \\
    \cline{2-9}
                         & $\mu^V$              & \tcb{0.86}               & \tcb{0.74} & 0.71 & \tcb{0.68}      & 0.92 & 0.90 & 0.91      \\
    \hline
    \end{tabular}
}
\end{table}

With the temperature increase, the average test score for manufacturer identification almost retain the same score as of the nominal temperature. However, the average part-number identification across all manufacturers is slightly degraded (presented in red in Table \ref{Tab:AccHT}); for example, the $F_{1}$ score to identify the target class reduced from 0.71 to 0.68 (presented in red in Table \ref{Tab:AccHT}). Especially, the SRAM chips from IDT and REA are affected most while we have increased the temperature. For IDT, the average $F_{1}$ score for part-number identification is reduced by 19\% (0.53 to 0.43), and for REA, the $F_{1}$ score is degraded by 9\%. For IDT and REA, we expected such results as the features associated with those part-numbers are very closely distributed (as explained in the previously). Hence, a slight thermal noise on start-up data impacted the corresponding classifiers heavily. Interestingly, the classification score improved by a little margin for AMI, although chips from AMI1 and AMI2 are closely located in feature-space (Fig. \ref{fig:spec_detection_AMI}). With closer observation, we have found that the features from AMI1 impacted heavily at higher temperatures and shifted away from the AMI2, which provided a relatively better separation between AMI1 and AMI2. The temperature sensitivity of AMI1 is not surprising as AMI1 possesses a lower temperature grade than AMI2.

In Table \ref{Tab:AccM}, \ref{Tab:AccP} and \ref{Tab:AccHT}, we trained the classifier using only one entropy source (i.e., all features are extracted from start-up data at nominal voltage). Our proposed technique can be further improved if more features can be extracted from different entropy sources. For example, we collected three sets of start-up data at low voltage (3.0V), nominal voltage (3.3V), and high voltage (3.6V) from all IDT chips. Then, we only extracted feature $\Phi_1$, $\Phi_4$, $\Phi_6$, and $\Phi_7$ from all of those three datasets and concatenated them in a single feature set (total 12 features). We trained ML models from train samples as we have done earlier and used the model to identify part-numbers from IDT. The outcome of the experiment was aligned with our expectation; The average $F_{1}$ score of part-number identification is improved to 0.6 from 0.53 (presented in cyan in Table \ref{Tab:AccP}).

\subsection{Identifying Recycled Memory Chips} \label{sec:RecycleSRAM}
As we explained in Sec. \ref{sec:method}, the recycled (aged) and fresh (aged) SRAM chip can be distinguished by only observing the number of 1's in start-up data \cite{Recycled:Guin}. As our method also uses the number of 1's as a feature ($\Phi_1$), our method is more generalized. Moreover, identifying the recycled chips by observing the number of 1's is only possible if the SRAM chips experience more logic ``1" than the logic ``0" (skewed data distribution). Although such a scenario is practical over the natural usage of the SRAM chips, we conducted an experiment without making the assumption of skewed data distribution.
\begin{figure*}[ht!]
\centering
\begin{subfigure}[]{0.48\textwidth}
    \includegraphics[width=1\textwidth, trim=1.5cm 0.2cm 0.8cm 0.5cm, clip]{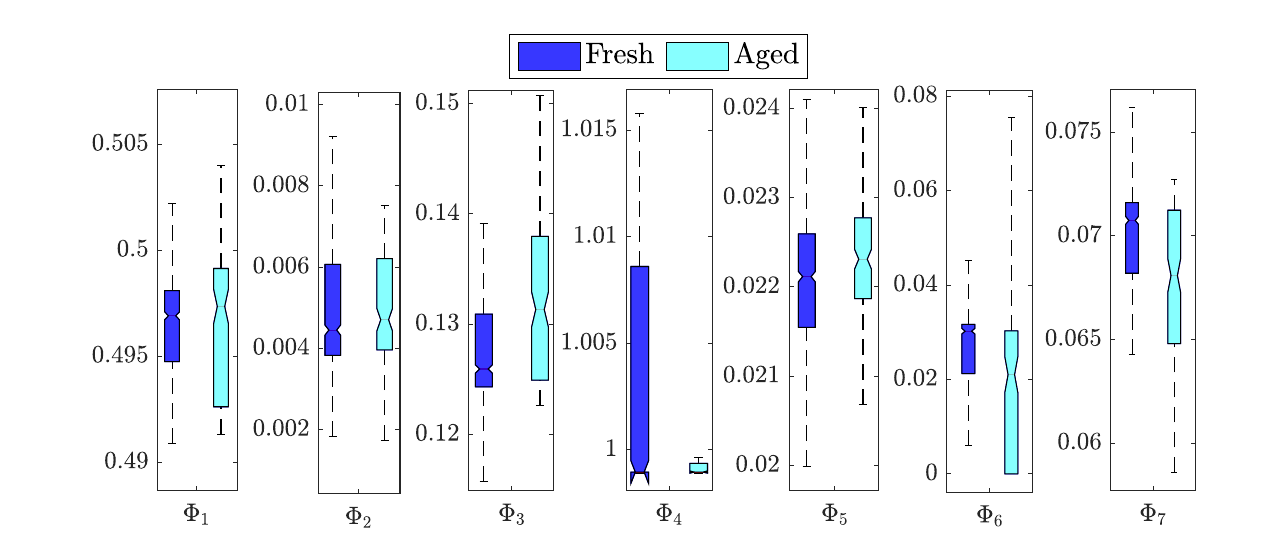}
    \caption{Cypress Semiconductor (CY)}
    \label{fig:CY_freshVSrecycled}
\end{subfigure}
\begin{subfigure}[]{0.48\textwidth}
    \includegraphics[width=1\textwidth, trim=1.4cm 0.3cm 1.1cm 0.5cm, clip]{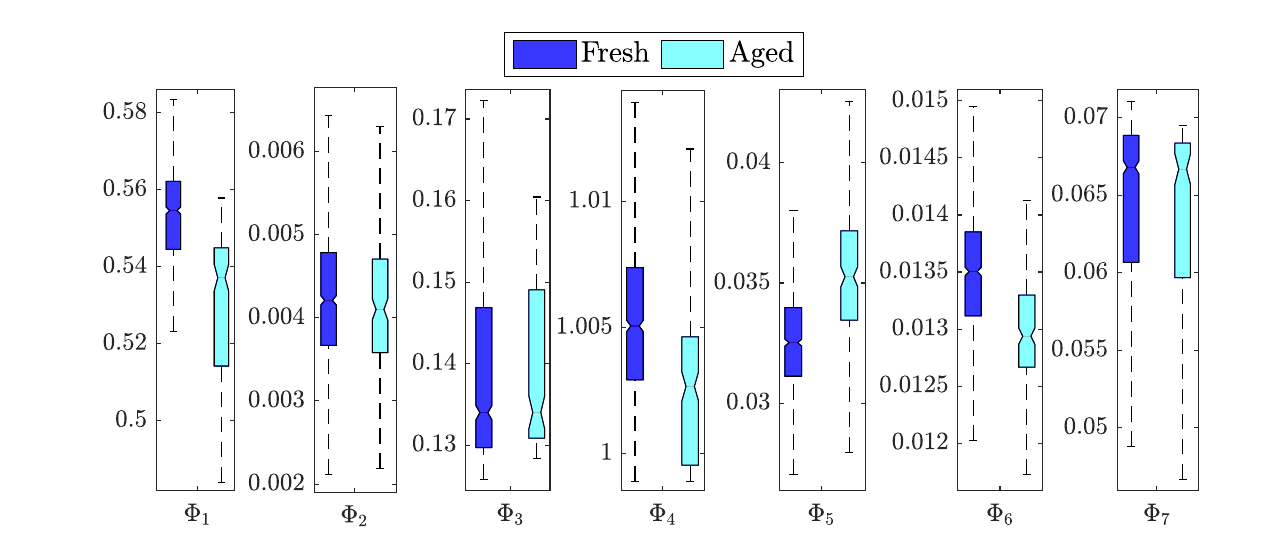}
    \caption{Integrated Device Technology (IDT)}
    \label{fig:IDT_freshVSrecycled}
\end{subfigure}
\caption{Visualizing feature distribution: fresh vs. aged.}
\label{Fig:freshVSrecycled}
\end{figure*}

Our experiment used the ``accelerated aging" \cite{SCARE_GUO} method by continuously writing random bits on SRAM chips. In accelerated aging, we exposed the memory chips in high voltage (3.6V) and high temperature ($80^{\circ}$C) for 1 hour and continuously wrote different random numbers (with the normal distribution of 0's and 1's). The temperature of the chip was controlled by a thermostream system \cite{thermostream}. We collected start-up data before and after the aging process and extracted features from them. The aging process is time-consuming, and we had limited access to the thermostream system; hence, we were only able to experiment with a limited number of chips. Our experiment used used 2 SRAM chips from each part-number of CY and IDT (10 CY chips and 12 from IDT). Although this small number of chips is not sufficient for the ML algorithm, our experiment demonstrates the impact of the aging process on features that are selected in Sec. \ref{sec:FeatSelect}.

We presented the distribution of the features from fresh chips and aged chips in Fig. \ref{Fig:freshVSrecycled}. Fig. \ref{fig:CY_freshVSrecycled} and \ref{fig:IDT_freshVSrecycled} represent feature distribution for CY and IDT, respectively. Because of using random numbers (uniform distribution of 0's and 1's) to age the device, we have an unpredictable shift on the $\Phi_1$ (number of 1's) distribution, which is used in previously proposed method to identify recycled SRAM chips \cite{Recycled:Guin}. However, we observed some other features might be extremely useful even with the presence of the uniform data pattern. For example, the distribution of $\Phi_6$ (number of noisy signature bits) always tends to shift towards 0. With sufficient aging, the distribution of $\Phi_6$ from the fresh and aged chips will be completely separable. During the aging process with the random data pattern, the number of 0's or 1's experienced by each memory cell will be a normal distribution. Hence, some of the noisy signature bits (located at distribution tail) will experience more 0's or 1's than others. With the same argument presented in \cite{Recycled:Guin}, we can argue that this will bias those noisy signature bits either toward ``1" or ``0" and reduce the total number of noisy signature bits (see Sec. \ref{sec:distFactor} for details). Note that, even with the biased data pattern (dominate by ``0" or ``1"), the number of the noisy signature bits will also be reduced (noisy signature bits will achieve either stable ``1" or ``0"). 

We also observe a shift in the distribution of other features. For example, now the compression ratio is closer to 1 (distribution of $\Phi_4$). This is also understandable as the random distribution on the data pattern biased the SRAM cells randomly and randomizes the start-up data. However, this distribution might shift upward if the usage data pattern is biased towards either ``0" or ``1" (i.e., start-up data will have more ``1" or ``0" after usages). Hence, imposing a boundary condition on $\Phi_4$ distribution might also be helpful to identify recycled SRAMs.

\subsection{Evaluation Time} \label{sec:EvalTime}
Our proposed method is aimed to identify counterfeit memory chips from the consumer end (or at least start-up signature should be collected at consumers' end (See Fig. \ref{fig:Protocol})). Nevertheless, our proposed method can also be scaled up for bulk testing. A single FPGA or high-speed embedded system can be used to collect and analyze data for bulk testing purposes. The average access time for a Commercial off-the-shelf (COTS) SRAM is $<$15ns/word. Hence the total access time for a 4Mb (256K$\times$16) SRAM is $<$4ms ($\approx$15ns$\times$256K). In our experiment, we have collected start-up data 20 times. Additionally, to avoid the discharge inversion effect, the sampling interval of 10s should be more than sufficient \cite{Liao:dischargeInversion}. The inference time of the machine learning model is very negligible compared to the data collection process (order of $\mu$s). Hence, the total time required to test an SRAM chips' authenticity is $\sim$3min ($\approx$19$\times$10s$+$20$\times$4ms), which is the time required for collecting the SRAM start-up data.

\section{Discussion} \label{sec:discussion}
\subsection{Scope and Limitations} \label{subsec:scope_lim}
Identifying memory manufacturer and part-number are useful for identifying many counterfeitings, which might be introduced at a different supply chain stage. In our proposed method, we extracted a set of features to capture the architectural, layout, and process variations among the memory chips manufactured by different manufacturers with different specification sets. Although identifying manufacturer/part-number does not identify recycled chips directly, our proposed feature-set can differentiate between new and recycled chips, as the device properties are changed over time (explained in Sec \ref{sec:distFactor} and \ref{sec:RecycleSRAM}). The tampered and out-of-spec/defective memory chips usually have some fundamental differences at the silicon level, either intentionally introduced by the untrusted facility center or due to the fabrication imperfection. Therefore, feature-set extracted from these types of counterfeit chips should have different characteristics from the authentic chips. For reverse-engineered chips, the counterfeiter usually recovers the functional netlist by depackaging the chip by some electrochemical process and inspecting the chip die by some imaging techniques \cite{Forte:CHES, CounterfeitIC:UGuin}. Once the netlist is constructed, the counterfeiter can use it for layout design and fabricating new chips without incurring any R\&D cost on developing the netlist. The reverse-engineered memory chips are usually differed by layout design and process variation. Therefore, our proposed feature set should be able to capture this type of counterfeiting. In cloned counterfeit type, the counterfeit chips are at least differed by the process variation, i.e., the final GDSII is cloned by the counterfeiter but fabricated in a different fabrication facility. Therefore, the cloned chips can also be identified by our proposed technique. Finally, the remarked chips are completely different from the authentic chips, where the manufacturer name and the part-number are altered; therefore, our proposed method can also identify the cloned chips. Unfortunately, our proposed method might not be able to identify the forged documented and overproduced chips as they are usually designed and fabricated with the same entity; hence, they usually have similar architectural, layout, and process variations. 

\subsection{Future Work} \label{sec:futureWork}
In our future work, we aim to explore more robust entropy sources across the temperature and voltage variation but are sensitive to usage. Additionally, ML model accuracy largely depends on feature selection/extraction techniques; hence, we emphasize exploring more features to improve our algorithm. For instance, many well-known features that work well with the binary image classification \cite{Heurtier:imageFeature} might also be used to extract features from binary memory signature. We also like to explore the correlation between the feature-set and the technology node, which might provide some deeper insight into features that can add value to our feature selection technique.

\section{Conclusion}\label{sec:conclusion}
This article presents a non-invasive and low-cost technique to (i) identify the memory manufacturer and part-number and (ii) recycled SRAM chips without requiring any additional hardware. This proposed framework has potential to use for other volatile and nonvolatile memory chips and help stop  spreading them in the supply chain. Finally, to train a more practical and accurate ML model, we need more train samples which might require an industry scale setup and crowd-sourcing.

\begin{acks}
This work was supported in part by the National Science Foundation under Grant Number CNS-1850241.
\end{acks}

\bibliographystyle{ACM-Reference-Format}
\bibliography{ref}










\end{document}